\documentclass[useAMS,usenatbib]{mn2e}

\usepackage{graphicx}
\usepackage{mathrsfs}
\usepackage{natbib}
\usepackage{amssymb}
\usepackage{amsmath}
\usepackage{mathtools}
\usepackage{xcolor}
\usepackage{txfonts}
\usepackage{ulem}
\DeclareSymbolFont{cmletters}{OML}{cmm}{m}{it}
\DeclareMathSymbol{v}{\mathalpha}{cmletters}{"76}
\bibpunct[,]{(}{)}{;}{a}{}{,}

\def\totd{{\mathrm{d}}}
\def\sun{{\odot}}

\title[GRMHD evolution of NS merger disks]{Long-term GRMHD Simulations
of Neutron Star Merger Accretion Disks: Implications for Electromagnetic Counterparts}
\author[Fern\'andez, Tchekhovskoy, Quataert, Foucart, \& Kasen]{
Rodrigo Fern\'andez$^{1}\thanks{E-mail: rafernan@ualberta.ca}$, Alexander Tchekhovskoy$^{2}$,
Eliot Quataert$^{3,4}$, Francois Foucart$^{5}$, 
\newauthor $\phantom{A}$and Daniel Kasen$^{3,4,6}$\\
$^1$ Department of Physics, University of Alberta, Edmonton, AB T6G 2E1, Canada\\
$^2$ Center for Interdisciplinary Exploration \& Research in Astrophysics (CIERA),
Physics \& Astronomy, Northwestern University, Evanston, IL 60202, USA\\
$^3$ Department of Astronomy \& Theoretical Astrophysics Center, University of California, Berkeley, CA 94720, USA\\
$^4$ Department of Physics, University of California, Berkeley, CA 94720, USA\\
$^5$ Department of Physics and Astronomy, University of New Hampshire, Durham, NH 03824, USA\\
$^6$ Nuclear Science Division, Lawrence Berkeley National Laboratory, Berkeley, CA 94720, USA\\
}

\begin{document}

\date{Submitted to MNRAS}
\pagerange{\pageref{firstpage}--\pageref{lastpage}} 
\pubyear{2014}
\maketitle
\label{firstpage}

\begin{abstract}
We investigate
the long-term evolution of black hole accretion disks formed
in neutron star mergers. These disks expel matter that 
contributes to an $r$-process kilonova, and can produce relativistic jets 
powering short gamma-ray bursts. Here we report the results of a three-dimensional, 
general-relativistic magnetohydrodynamic (GRMHD) simulation of such a disk which is evolved
for long enough ($\sim 9$\,s, or $\sim 6\times 10^5 r_{\rm g}/c$) to achieve completion of
mass ejection far from the disk. Our model starts with a poloidal field, and 
fully resolves the most unstable mode of the magnetorotational instability.
We parameterize the dominant microphysics and neutrino cooling effects, and compare with axisymmetric 
hydrodynamic models with shear viscosity. 
The GRMHD model ejects mass in two ways: a prompt MHD-mediated outflow and a late-time,
thermally-driven wind once the disk becomes advective. The total amount of unbound
mass ejected ($0.013M_\odot$, or $\simeq 40\%$ of the initial torus mass) is twice as much as in
hydrodynamic models, with higher average velocity ($0.1c$) and a broad electron fraction
distribution with a lower average value ($0.16$).
Scaling the ejected fractions to a disk mass of $\sim 0.1M_\odot$ can account for the red 
kilonova from GW170817 but underpredicts the blue component.
About $\sim 10^{-3}M_\odot$ of material should undergo neutron freezout and could
produce a bright kilonova precursor in the first few hours after the merger. 
With our idealized initial magnetic field configuration, we obtain a robust jet and
sufficient ejecta with Lorentz factor $\sim 1-10$ to (over)produce the non-thermal emission 
from GW1708107.
\end{abstract}


\maketitle

\section{Introduction}

The recent detection of the neutron star (NS) merger 
GW170817\footnote{Also known as GRB170817A, SSS17a, AT 2017gfo, and DLT17ck} 
in gravitational- and electromagnetic waves (\citealt{ligo_gw170817_gw,ligo_gw170817_multi-messenger}, and
references therein) has advanced several outstanding issues in astrophysics. 
It has established neutron star mergers as an important (if not dominant) site
of $r$-process element production 
(e.g., \citealt{kasen_2017,cote_2018, hotokezaka_2018}), 
provided unambiguous association between
a neutron star merger and a short gamma-ray burst \citep{ligo_gw170817_grb}, and 
set constraints on the dense-matter equation of state 
(e.g., \citealt{bauswein_2017,margalit_2017,rezzolla_2018,chatziiouannou_2018,raithel_2018,de_2018,ligo_gw170817_eos}).

Evidence for the $r$-process comes from the photometric and spectroscopic
properties of the  observed kilonova (e.g., \citealt{cowperthwaite_2017,chornock_2017,
drout_2017,tanaka_2017,tanvir_2017}). This type of transient had been predicted to arise out 
of sub-relativistic,
neutron-rich ejecta from the merger that is radioactively heated by freshly produced
$r$-process elements \citep{Li&Pacyznski98,Metzger+10b,Roberts+11,tanaka2016,metzger_2017}. 
The optical opacity of lanthanides and actinides ($A>130$, produced
by the $r$-process) is such that the transient was expected to evolve from blue optical to near
infrared within a few days \citep{Kasen+13,tanaka2013,Barnes&Kasen13,fontes2015}, as observed. 
Also, the temporal evolution of the bolometric luminosity is consistent with the time-dependence 
of the radioactive heating rate from the $r$-process (e.g., \citealt{rosswog_2017}).

Two main mass ejection channels operate in neutron star mergers: dynamical ejecta and outflows from
the remnant accretion disk. The former is launched on the dynamical time of the merger ($\sim$\,ms) by
tidal forces and hydrodynamic interactions (e.g., \citealt{bauswein2013,Hotokezaka+13}).
Numerical relativity simulations predict this material to be sufficiently
neutron-rich to produce mainly $A > 130$ elements, with varying amounts of lighter material 
depending on the equation of state (EOS) of dense matter and the treatment 
of neutrino physics 
(e.g., \citealt{wanajo2014,roberts2017,radice2016,foucart2016a,foucart2016b}).
While magnetic fields are not expected to significantly alter the dynamics of the merger
(e.g., \citealt{endrizzi_2016}) they can lead to some mass ejection on the dynamical 
time (e.g., \citealt{kiuchi2014,shibata_2017}).
For the particular case of GW710817, the amount of dynamical ejecta
expected is smaller than the total $r$-process mass inferred from the 
kilonova (e.g., \citealt{ligo_gw170817_ejecta,shibata_2017b}; however see 
\citealt{kawaguchi_2018} for a different kilonova mass estimate).

The remnant accretion disk evolves on longer timescales ($\sim 100$~ms - $10$~s) 
and ejects mass through a combination of physical processes (see, e.g., \citealt{FM16} for an overview).
Immediately after the merger, the disk is sufficiently hot and dense for neutrinos
to be the primary cooling channel, with most of the nuclei fully dissociated into 
nucleons \citep{popham1999,ruffert1999,Narayan+01,Chen&Beloborodov07}.
A key property of these disks is that they transition to being fully advective once the density 
drops and weak interactions freeze-out on a timescale of $\sim 300$\,ms to $1$\,s, making
them prone to launching outflows \citep{Metzger+09a}.

Our current understanding of the long-term disk evolution is based primarily 
on axisymmetric hydrodynamic simulations
that include the required microphysics and neutrino treatment at various levels of
sophistication, but which model angular momentum transport through an imposed
shear stress with parameterized viscosity 
(e.g., \citealt{FM13,Just+15,fujibayashi_2018}).
With this physics included, the outflow is driven primarily by viscous heating and nuclear
recombination, with neutrino heating being sub-dominant when the central object is a black hole (BH).
The amount of mass ejected in these simulations lies in the range $\sim 5-20\%$ of the initial disk mass
after an evolution time of $\sim 10$\,s, with quantitative details depending primarily on the 
properties of the central object (a much larger fraction of the disk can be ejected if a 
hypermassive neutron star forms; \citealt{MF14}). The composition of the outflow involves 
mainly light $r$-process elements, with varying amounts of material with $A > 130$ depending on 
parameters such as the strength of angular momentum transport or the lifetime of a hypermassive neutron 
star (HMNS) \citep{Just+15,martin2015,wu2016,lippuner_2017}.

It is generally accepted, however, that angular momentum transport in astrophysical accretion disks
operates via magnetohydrodynamic (MHD) turbulence driven by the  magnetorotational 
instability (MRI; \citealt{balbus1991}). Early GRMHD models of NS-NS/BH-NS merger remnant disks 
employed axisymmetric (2D) simulations that start with an initially poloidal field geometry, 
and which include the relevant neutrino processes, but  (1) had too high an ambient density to 
allow for a significant outflow and/or (2) did not evolve the system 
for long enough to achieve the radiatively-inefficient state and 
{completion of} mass ejection \citep{shibata2007,shibata2012,janiuk2013,janiuk_2017}. 
In addition, it is well-known that in axisymmetry, 
as a consequence of the anti-dynamo theorem~\citep{cowling33}, MRI turbulence dissipates 
within $\sim 10$ disk orbits (e.g. \citealt{haw00}) and hence angular momentum transport
cannot be sustained for the required timescales. 

Recently, \citet{siegel_2017a,siegel_2018} have reported the first three-dimensional (3D)
GRMHD simulation of an accretion disk around a black hole remnant. The simulation uses a physical
equation of state that includes recombination of nucleons into alpha particles, and 
accounts for neutrino cooling via a leakage scheme. They start their simulation with
an equilibrium torus and an initial poloidal field, evolving the disk for $\sim 400$\,ms.
Strong outflows are obtained, and by the end of their simulation 20\% of the initial disk mass is ejected
as unbound matter at a radius of $10^8$\,cm, with 60\% of the disk mass accreted. 
Since by that time accretion onto the BH is mostly complete, they surmise that the remaining 20\% would 
continue to be ejected as an unbound outflow if the simulation was continued to longer times. 
\citet{nouri_2017} also
studied the GRMHD evolution of an accretion disk mapped from a 3D non-magnetized numerical
relativity simulation of a BH-NS merger, endowing the disk with a poloidal field and 
following its evolution for $60$\,ms, 
reaching a fully developed MRI and comparing with the non-magnetized case.

While there is an existing body of work on the long-term evolution of black hole accretion disks, 
including a number of 3D GRMHD studies (e.g., \citealt{mckinney2012}), work has 
focused primarily on systems arising in X-ray binaries and active galactic nuclei, 
for which thermodynamic conditions, disk size, and the effect of photons and neutrinos are 
very different than for NS-NS/NS-BH mergers. Because of these differences, the 
results of previous work that focused on the sub-relativistic disk outflow 
(e.g., \citealt{narayan2012,sadowski2013}) are not directly applicable to the merger problem . 

In this paper we close the gap in disk evolution time by
performing long-term GRMHD simulations of NS merger accretion disks that 
for the first time achieve {completion of} mass ejection (i.e., most of the initial disk material either accreted or ejected). 
In order to minimize the computational cost and evolve our simulations for as long as possible, 
we employ a number of approximations to the microphysics and neutrino treatment.
We focus on understanding the basic properties of the sub-relativistic outflow when MHD 
turbulence transports angular momentum.
To carry out our simulations, we extend the
GRMHD code {\tt HARMPI}\footnote{Available at
  https://github.com/atchekho/harmpi \label{fn:harmpi}} to include the dominant microphysics and neutrino source terms. 
Simulations are evolved for long enough (several times $10^5 r_g/c$, with $r_g = GM_{\rm bh}/c^2$ the gravitational 
radius of a black hole of mass $M_{\rm bh}$) to achieve the advective state
in the disk evolution and to reach {completion of} mass ejection. While our models resolve the MRI, 
we consider our work to be exploratory in nature, because not only more spatial 
resolution but also more physics and realistic initial conditions are 
required to make quantitative predictions on the wind contribution to the kilonova and $r$-process nucleosynthesis,
particularly due to the sensitivity of the latter to the exact outflow composition.

{Given that until now only hydrodynamic disk models have been evolved into the advective state,
and that a key mass ejection mechanism (thermal energy deposition by angular momentum 
transport and nuclear recombination) is present in both approaches, we compare the results of
our GRMHD simulation with those from hydrodynamic models that employ an alpha viscosity to 
transport angular momentum.
The goal is to identify similarities and differences in mass ejection, thus providing a 
solid foundation to understand the behavior of the GRMHD model.}

The structure of the paper is the following. Section~\ref{s:methods} describes
our computational setup, \S\ref{s:results} presents
our results, \S\ref{s:obs_implications} discusses the observational
implications, and \S\ref{s:summary} closes with a summary and future prospects.
The Appendix presents a detailed description of the microphysics included, and a comparison with 
more complete hydrodynamic models.

\section{Computational Methodology}
\label{s:methods}

\subsection{Physical Model}
\label{s:initial_model}

Our goal is to study the long-term evolution of the accretion disk that forms 
during the dynamical phase of a NS-NS or a BH-NS merger, assuming
that the central remnant is a promptly-formed BH. This requires
the mass of the remnant to exceed a certain threshold that depends on the 
dense matter EOS (e.g., \citealt{shibata2006}).
Since the spacetime around the BH settles into a stationary configuration a few light crossing times 
after the end of the merger (e.g., \citealt{lehner2014}), 
we adopt a static Kerr metric. We adopt a black hole mass $M_{\rm bh} = 3M_\odot$ with 
spin $a=0.8$ {(e.g., \citealt{Kastaun_2013,Bernuzzi_2014})} 
as a fiducial value for a generic NS-NS merger\footnote{At the time our 
parameters were chosen and our runs were completed, GW170817 had not yet been detected.}.
Disks formed in BH-NS mergers are expected to follow the same qualitative evolution
as disks from NS-NS mergers, with quantitative differences due to the 
higher mass of the central BH.

In order to maximize the physical time evolved in our simulation given
fixed computational resources and {\tt HARMPI} capabilities, we employ
an EOS with constant adiabatic index $\gamma_{\rm ad}$ in all of our calculations. 
We choose the value of $\gamma_{\rm ad}$ by
comparing the results of 2D hydrodynamic wind models that
use a physical EOS \citep[ hereafter F15]{FKMQ14} with those obtained using 
an ideal gas (\S\ref{s:hydro_methods}). 
The results of this comparison are presented in Appendix~\ref{s:microphysics}.
Good agreement in the wind properties is obtained when using
$\gamma_{\rm ad}=4/3$, which is consistent with the dominance of radiation
pressure at large radii. Minor quantitative differences are obtained 
in the inner regions of the disk, where the pressure at early times 
is dominated by an ideal gas of non-relativistic ions, and where electrons 
are relativistic and mildly degenerate.

The temperature of the gas is obtained by assuming that
the total pressure $P$ has contributions from radiation and
an ideal gas of neutrons, protons, and electrons
\begin{eqnarray}
P & = & P_{\rm gas} + P_{\rm rad}\\
\label{eq:ptot_equation}
P & = & \left[1 + Y_e \right] \frac{\rho k T}{m_n} + \frac{1}{3}a_{\rm rad}T^4,
\end{eqnarray}
where $\rho$, $T$, and $Y_e$ are the gas density, temperature, and electron fraction, 
$a_{\rm rad}$ is the radiation constant, and we have neglected
the mass difference between neutrons and protons. In the densest
parts of the disk, at early times, the pressure is dominated by
nucleons, so the assumption of non-degeneracy for the electron pressure 
contribution is not important. As shown in \S\ref{s:adiabatic_index}, at late
times the wind is dominated by radiation pressure, thus the
assumption of non-degeneracy of electrons in equation~(\ref{eq:ptot_equation}) 
is also secondary. Electron degeneracy is nevertheless included in the
calculation of the neutrino source terms below.
Equation~(\ref{eq:ptot_equation}) yields the temperature given
the pressure, density, and electron fraction. 

We account for the emission of electron neutrinos and antineutrinos
through charged-current weak interactions on nucleons. Neutrino
absorption is ignored, which is a reasonable first approximation when
a black hole forms promptly (e.g., \citealt{Just+15}).
We employ the approximate emission rates of \citet{janka2001}, which
are valid for temperatures $kT \gg (m_n-m_p)c^2 \simeq 1.3$~MeV. 
At lower temperatures, neutrino emission is unimportant, so this
approximation captures the dominant effect. Neutrino and antineutrino
emission causes the electron fraction and internal energy to change in the local fluid
frame at rates that are, respectively,
\begin{eqnarray}
\label{eq:Gnu}
\Gamma & = & \frac{4c(1+g_A^2)}{(hc)^3}\mathcal{F}_4(0)\left[\frac{G_{\rm F}}{(\hbar c)^2} \right]^2
	   (kT)^5 D_4(\eta_e,X_n,X_p) \nonumber\\
        & \simeq & 0.22\, T_{10}^5\, D_4(\eta_e,X_n,X_p) \quad[\textrm{s}^{-1}]\\
\label{eq:Qnu}
Q & = & - \frac{4c(1+g_A^2)}{m_n (hc)^3}\mathcal{F}_5(0)\left[\frac{G_{\rm F}}{(\hbar c)^2} \right]^2
	   (kT)^6 D_5(\eta_e,X_n,X_p) \nonumber\\
        & \simeq & -8.9\times 10^{17}\, T_{10}^6\, D_5(\eta_e,X_n,X_p) \quad[\textrm{erg g}^{-1}\textrm{ s}^{-1}],
\end{eqnarray}
where $X_n$ and $X_p$ are the mass fractions of neutrons and protons, respectively,
and the degeneracy parameter is $\eta_e = \mu_e/(kT)$, with $\mu_e$ the electron
chemical potential (Appendix~\ref{s:microphysics}). Other physical constants have their standard meanings.
Suppression of emission in neutrino-opaque regions is approximated by multiplying 
$\Gamma$ and $Q$ by $e^{-\tau_\nu}$, with
\begin{equation}
\tau_\nu = \rho / 10^{11}\textrm{g cm}^{-3}
\end{equation} 
an approximation to the neutrino optical depth to charged-current weak interactions. This approximation
is acceptable as long as the optical depth in the disk is small (see \S\ref{s:initial_conditions}
for details on the initial torus properties).

The effects of degeneracy and composition enter into the neutrino source terms (equations~\ref{eq:Gnu}-\ref{eq:Qnu}) 
through the dimensionless functions
\begin{align}
\label{eq:D4}
D_4(\eta_e,X_n,X_p) &= \left[X_n\mathcal{F}_4(-\eta_e)-X_p\mathcal{F}_4(\eta_e)\right]/\mathcal{F}_4(0)\\
\label{eq:D5}
D_5(\eta_e,X_n,X_p) &= \left[X_n\mathcal{F}_5(-\eta_e)+X_p\mathcal{F}_5(\eta_e)\right]/\mathcal{F}_5(0),
\end{align}
where
\begin{equation}
\mathcal{F}_j(\eta) = \int_0^\infty \frac{x^j\,\totd x}{e^{x-\eta}+1}.
\end{equation}
are the Fermi functions. Analytical approximations are adopted for these integrals, with details
provided in Appendix~\ref{s:microphysics}.

The equilibrium electron fraction implied by equation~(\ref{eq:Gnu}), including only
neutrons and protons, is given by the condition $D_4(\eta_e,1-Y_e^{\rm eq},Y_e^{\rm eq}) = 0$, or
\begin{equation}
\label{eq:ye_eq}
Y_e^{\rm eq} = \frac{\mathcal{F}_4(-\eta_e)}{\mathcal{F}_4(\eta_e)+\mathcal{F}_4(-\eta_e)} \simeq
\begin{dcases}
\frac{1}{2(1 + \eta_e)} & \eta_e \ll 1\\
\noalign{\smallskip}
\frac{5!}{\eta_e^5}e^{-\eta_e}               & \eta_e \gg 1.
\end{dcases}
\end{equation}

We include the contribution of the nuclear binding energy of $\alpha$ particles 
to the internal energy through a source term
\begin{eqnarray}
\label{eq:qalpha}
\Delta q_{\rm nuc} & = & \frac{B_\alpha}{m_\alpha}\,\Delta X_\alpha\\
         & = & 6.8\times 10^{18}\,\Delta X_\alpha\,\textrm{erg g}^{-1},
\end{eqnarray}
where $B_\alpha = 28.3$~MeV and $m_\alpha$ are the nuclear binding energy 
and mass of an $\alpha$ particle, respectively, and $\Delta X_\alpha$ is the
change in the mass fraction of alpha particles ($X_\alpha$).
The source term is applied implicitly at each time step in the fluid rest frame,
\begin{equation}
\label{eq:implicit_energy_update}
e_{\rm int}^{n+1} - \frac{B_\alpha}{m_\alpha}X_\alpha^{n+1} = e_{\rm int}^n - \frac{B_\alpha}{m_\alpha}X_\alpha^{n} + Q^n\Delta t,
\end{equation}
where $e_{\rm int}$ is the specific internal energy, 
$\Delta t$ is the simulation time step,
and the time discretization is labeled by the superscript $n$ (an additional factor of $(u^t)^{-1}$ is added
to $Q$ in GRMHD, c.f. equation~\ref{eq:energy_update_gr}). 
With this notation, $\Delta X_\alpha = X_\alpha^{n+1}-X_\alpha^{n}$.
Equation~(\ref{eq:implicit_energy_update}) is applied
after $Y_e$ has been updated with the neutrino source term in equation~(\ref{eq:Gnu}), 
hence $\rho$ and $Y_e$ are kept constant in the iteration
(a simultaneous update of the temperature via equation~[\ref{eq:ptot_equation}] is also required).
The mass fraction $X_\alpha$ is computed using the analytic fit to Nuclear Statistical
Equilibrium (NSE) of \citet{Woosley&Baron92}\footnote{The fit of \citet{Woosley&Baron92} 
was derived in the context
of accretion-induced collapse of white dwarfs to neutron stars, for which $Y_e \simeq 0.5$.
For neutron-rich material, a fraction $(1-2Y_e)$ of the neutrons is
not available to form $\alpha$ particles.}
\begin{eqnarray}
\label{eq:xalpha_fit}
X_\alpha   & = & \mathrm{min}[2Y_e,2(1-Y_e)] (1 - \mathrm{min}[1,X_{\rm WB}])\\
\label{eq:WB92_fit}
X_{\rm WB} & = & 15.58\frac{T_{\rm MeV}^{9/8}}{\rho_{10}^{3/4}}\exp(-7.074/T_{\rm MeV}),
\end{eqnarray}
where $X_{\rm WB}$ is the sum of the mass fractions of neutrons and protons available
to make $\alpha$ particles.
The numerical prefactor in equation~(\ref{eq:WB92_fit}) has been adjusted slightly to improve 
agreement in the density regime of interest (Appendix~\ref{s:microphysics}).
The mass fractions of neutrons and protons are obtained for every new value of $X_\alpha$,
given $Y_e$, from conservation of baryon number and charge,
\begin{eqnarray}
\label{eq:mass_conservation_nse}
X_n + X_p + X_\alpha & = & 1\\
\label{eq:charge_conservation_nse}
X_p + \frac{1}{2}X_\alpha & = & Y_e.
\end{eqnarray}

\subsection{GRMHD Simulations in 3D}
\label{s:grmhd_methods}

To evolve the torus in GRMHD, we use {\tt HARMPI}$^{\ref{fn:harmpi}}$, an enhanced version of the serial open-source 
code {\tt HARM} \citep{gammie_2003,noble_2006}. Updates include extension
to three spatial dimensions and parallelization with the Message Passing Interface 
(Tchekhovskoy et al., in preparation). The code has also been extended to include
the physics described in \S\ref{s:initial_model}. Specifically, we solve the ideal
GRMHD equations on a Kerr metric, with energy source terms due to neutrinos and 
nuclear recombination, and supplemented by a lepton number conservation equation
for the evolution of $Y_e$:
\begin{eqnarray}
\label{eq:gr_mass_conservation}
(-g)^{-1/2}\,\partial_\mu\left(\sqrt{-g}\rho u^\mu\right) & = & 0\\
\partial_\mu (\sqrt{-g}T_\nu^\mu) & = &  \sqrt{-g}T_\lambda^k\Gamma^\lambda_{\nu k}\\
\partial_t (\sqrt{-g}B^i) & = & -\partial_j\left[ \sqrt{-g}(b^j u^i - b^i u^j)\right]\\
\label{eq:gr_lepton_conservation}
(-g)^{-1/2}\,\partial_\mu\left(\sqrt{-g}\rho \mathbf{X} u^\mu\right) & = & 0.
\end{eqnarray}
where $g = \det(g_{\mu\nu})$ is the determinant of the Kerr metric, $\rho$ is the rest mass 
density, $\mathbf{X}$ is a composition vector that includes $Y_e$ and other mass fractions,
and $c=1$ has been assumed. As conventional, greek indices run in the range
$\{0-4\}$ while latin indices over $\{1-3\}$, with lower and upper indices denoting covariant
and contravariant components, respectively. 
The energy and lepton number source terms are applied
in the local fluid rest frame in between updates of conserved quantities,
\begin{eqnarray}
\label{eq:energy_update_gr}
\Delta \epsilon & = & \rho\left(Q\frac{\Delta t}{u^t} + \Delta q_{\rm nuc}\right)\\
\label{eq:ye_update_gr}
\Delta (\rho Y_e)      & = & \rho\Gamma \frac{\Delta t}{u^t}.
\end{eqnarray}
where $\epsilon = \rho e_{\rm int} = P/(\gamma_{\rm ad}-1)$ is the internal energy density,
and equation~(\ref{eq:energy_update_gr}) is solved implicitly at 
each time step (equation~\ref{eq:implicit_energy_update}).
In this unit system, $Q$ and $q_{\rm nuc}$ must be divided by $c^3/r_g$, and $\Gamma$ by $c/r_g$. 
The magnetic field 4-vector is denoted by $b^\mu$, and the magnetic field 3-vector is $B^i$.
The stress-energy tensor is 
\begin{equation}
T^{\mu \nu} = \left(\rho + \epsilon + P + b_\lambda b^\lambda\right)u^\mu u^\nu 
              + \left(P + \frac{1}{2}b_\lambda b^\lambda\right)g^{\mu\nu} -b^\mu b^\nu.
\end{equation}
With this notation, a factor of $(4\pi)^{-1/2}$ has been absorbed in the definition of $b$.

The code solves equations (\ref{eq:gr_mass_conservation})-(\ref{eq:gr_lepton_conservation}) 
in conservation form. Mapping from conserved to
primitive variables is carried out with the two-dimensional algorithm of \citet{noble_2006}.
Fluxes of conserved quantities are computed with a local Lax-Friedrichs method using
linear reconstruction and a monotonized central slope limiter. 
Enforcement of $\nabla\cdot\mathbf{B}=0$ is achieved with the
constrained transport method of \citet{toth_2000} that uses
a cell-centered representation of the magnetic field.

\begin{figure}
  \begin{center}
    \includegraphics*[width=0.9\columnwidth]{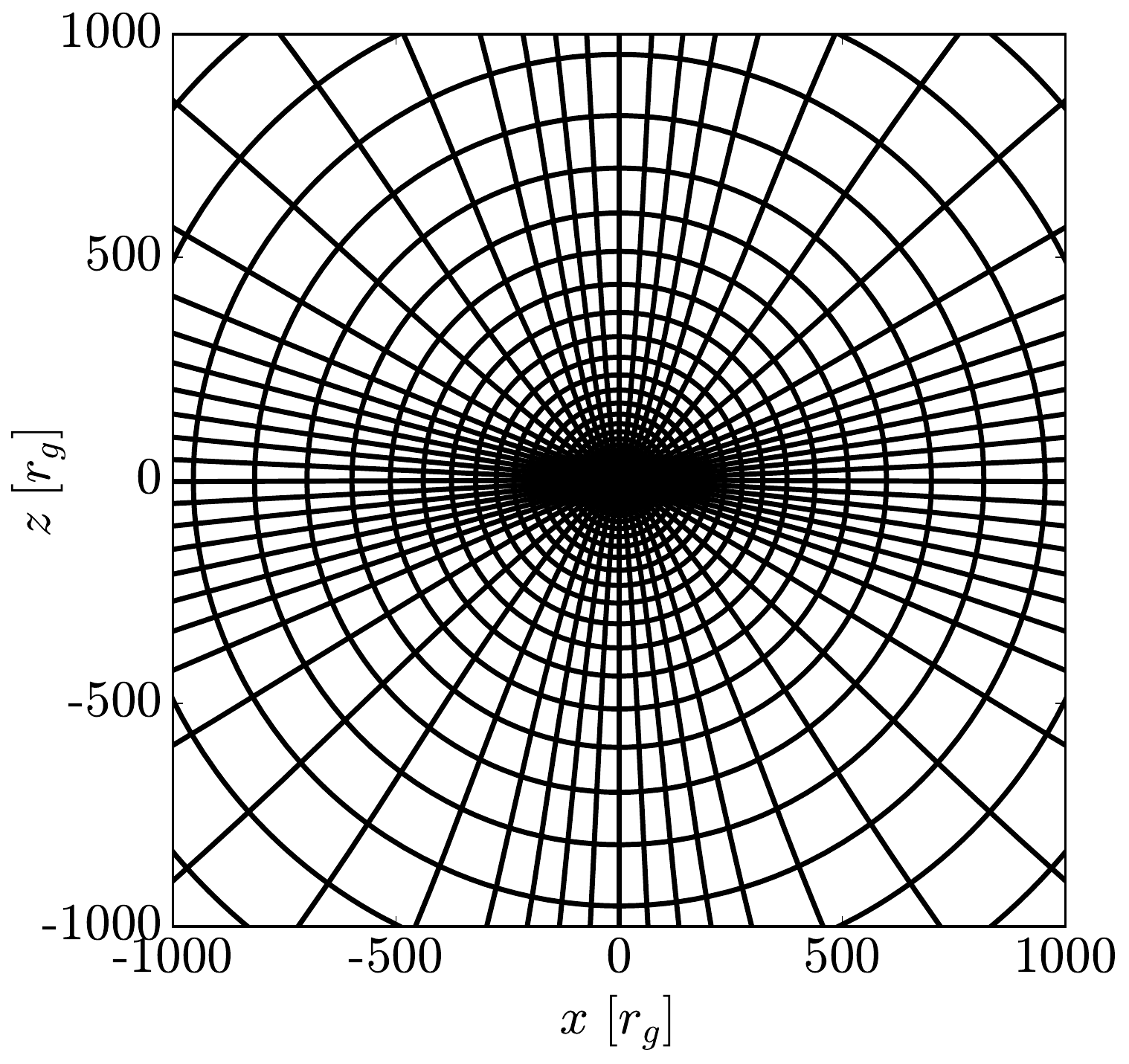}
  \end{center}
\caption{Meridional slice through the numerical grid used for our
  GRMHD simulation. The grid focuses the resolution on both the polar regions,
  where the relativistic jets form, and the equatorial plane, where
  the turbulent accretion disk resides. This allows us to obtain
  accurate descriptions of both the accretion disk, jets and outflows.
  For clarity, we show every 8th grid line.}
\label{f:grid}
\end{figure}

The computational grid is discretized in spherical-polar--like
modified Kerr-Schild coordinates.
We distribute the radial grid of $512$ cells as follows. The
first $458$ cells are logarithmically spaced 
between the inner radius $r_{\rm min}=0.87r_{\rm h} = 1.392r_g$
and the transition radius $r_{\rm tr}=10^4r_g$, yielding $\Delta r / r
\simeq 0.02$. Here, $r_{\rm h}=(1+\sqrt{1-a^2})r_g = 1.6r_g$ is the
event horizon radius. Thus, our inner radial boundary resides 7 cells
\emph{inside} of the event horizon, thereby ensuring that no signals
reach from the boundary to outside of the event horizon, as is
physically required.
To avoid the interaction between the expanding
accretion disk and the outer boundary, we extend the grid out to
$r_{\rm max} = 10^5r_g$: at $r>r_{\rm tr}$, the radial grid becomes
progressively sparse, with the remaining $54$ cells resolving the
distance between $r_{\rm tr}$ and $r_{\rm max}$.
The meridional grid of $256$ cells covers the range $[0,\pi]$, 
and is ``cylindrified'' close to the polar axis to save computational
time, as in \citet{sasha2011}. Figure~\ref{f:grid} shows that
the meridional grid focuses the resolution both in the
polar regions, to resolve the magnetized collimated relativistic jets,
and in the equatorial plane, to resolve the turbulent accretion
disk. This focusing approximately doubles the effective resolution in
the equatorial disk region, increasing it to $\approx510$ cells, and
increases the resolution by a
factor of $\approx 1.5$ in the polar regions, to $\approx370$ cells,
at large radii ($r\gtrsim 10^3r_g$).
The azimuthal grid is uniform, covering the range $[0,\pi]$ (half of the disk) with 64 cells.
The boundary conditions are outflow in the radial direction, reflective in the meridional
direction, and periodic in the azimuthal direction.

\subsection{Axisymmetric Hydrodynamic Simulations with shear viscosity}
\label{s:hydro_methods}

To compare with previous work, {and to help diagnose the GRMHD simulation,}
we also carry out 2D hydrodynamic simulations 
that transport angular momentum with a shear 
viscosity following the parameterization of \citet{shakura1973}. 
The models are carried out in a modified version of \textsc{FLASH3} 
\citep{fryxell00,dubey2009}
that has been equipped with the physics required to simulate
merger remnant accretion disks (\citealt{FM13,FM12,MF14}; F15).

The code solves the equations of mass, poloidal momentum, 
angular momentum, energy, and lepton number conservation in spherical polar
coordinates $(r,\theta)$ with source terms
due to gravity, shear viscosity, and neutrinos,
\begin{eqnarray}
\label{eq:mass_conservation}
\frac{\partial \rho}{\partial t} + \nabla \cdot (\rho\mathbf{v}_p) & = & 0\\
\label{eq:momentum_conservation}
\frac{\totd \mathbf{v}_p}{\totd t}  & = &
-\frac{1}{\rho}\nabla P  -\nabla\Phi_{\rm A} \\
\label{eq:angular_conservation}
\rho\frac{\totd \ell_z}{\totd t} & = & r\sin\theta\,(\nabla\cdot\mathbb{T})_\phi\\
\label{eq:energy_conservation}
\rho\frac{\totd e_{\rm int}}{\totd t} + p\nabla\cdot\mathbf{v}_p
& = & \frac{1}{\rho\nu}\mathbb{T}:\mathbb{T} + \rho \left(Q + \frac{B_{\alpha}}{m_\alpha}\dot X_\alpha\right)\\
\label{eq:lepton_conservation}
\frac{\totd Y_e}{\totd t} & = & \Gamma,
\end{eqnarray}
where $\totd/\totd t \equiv \partial/{\partial t} +\mathbf{v}_p\cdot\nabla$,
$\mathbf{v}_p$ is the poloidal velocity, $\ell_z = r\sin\theta v_\phi$
is the specific angular momentum along $z$. An explicit
form of the shear stress tensor $\mathbb{T}$ can be found in \citet{FM13}.
We only include the $r-\phi$ and $\theta-\phi$ components of the tensor
so that the shear due to orbital motion is subject to viscous stresses
but not poloidal motions such as thermally driven convection (following the approach 
of \citealt{stone1999}). 

The neutrino source terms and equation of state follow the
approximations described in \S\ref{s:initial_model}, while the gravity 
of the BH is modeled with the pseudo-Newtonian potential $\Phi_{\rm A}$ of \citet{artemova1996}. An
analytic form for this potential can be found in F15.

\begin{figure}
\includegraphics*[width=\columnwidth]{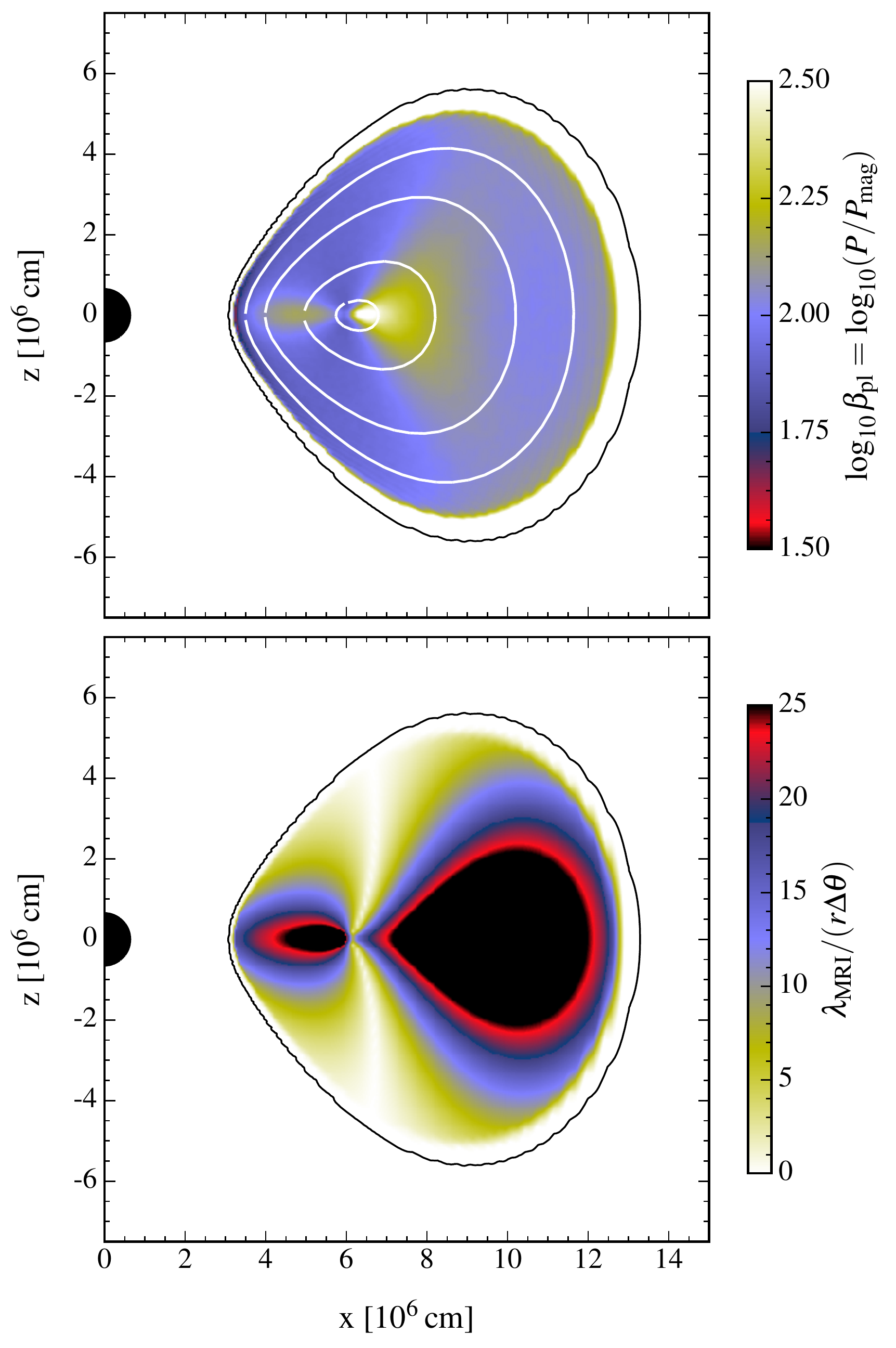}
\caption{Initial condition for the GRMHD model. The black circle marks the inner boundary of
the computational domain at $r_{\rm min} = 1.4r_g \simeq 6.2\times 10^5$\,cm. 
\emph{Top:} ratio $\beta_{\rm pl}$ of gas plus radiation pressure $P$ (equation~\ref{eq:ptot_equation}) 
to magnetic pressure $P_{\rm mag}$ (equation~\ref{eq:pmag_def}). 
The white contours show magnetic 
field lines, and the black contour shows the
isodensity surface $\rho = 10^6$\,g\,cm$^{-3}$, close to the edge of the disk. 
The maximum field strength is approximately $4\times 10^{14}$\,G.
\emph{Bottom:}
Number of meridional cells that resolve the wavelength $\lambda_{\rm MRI}$ 
of the most unstable mode of the poloidal MRI.
The black contour is the same as in the top panel.}
\label{f:initial_conditions}
\end{figure}

The computational grid covers the full range of polar angles $[0,\pi]$,
with a radial range that extends from the average between the horizon
and the ISCO radii on the inner side, $r_{\rm min} = (r_h + r_{\rm ISCO})/2\simeq 2.3r_g$, 
until a radius $10^4$ times larger. The spacing is logarithmic in radius and
uniform in $\cos\theta$. The baseline resolution is $128$ points
per decade in radius and $112$ points in the polar direction, yielding
approximately square cells at the equator with 
$\Delta r/r \simeq \Delta \theta \simeq 0.018 \simeq 1^\circ$.
The boundary conditions are set to outflow at both radial ends,
and reflecting in the $\theta$-direction.

\subsection{Initial Conditions}
\label{s:initial_conditions}

The accretion disk is formed out of gravitationally bound material that has 
enough angular momentum to resist direct collapse to the BH.
The material circularizes into a nearly axisymmetric configuration
a few orbital times after the merger ($\lesssim 10$~ms, \citealt{ruffert1999}). 
Given that the evolutionary timescales that we are interested in are much
longer than this circularization time, we take an equilibrium torus as 
the initial condition, and choose parameters to match representative merger systems.
The results of \citet{fernandez_2017} show that this approximation leads to
qualitatively the same results in the late time viscously-driven outflow as when
starting from the output of a dynamical merger calculation,
with quantitative modifications to the electron fraction mass distribution 
of the order of $\sim 10\%$. More realistic initial conditions for the electron
fraction are likely to affect material ejected promptly by magnetic fields (\S\ref{s:early_evolution}).
{Since this is the first long-term GRMHD simulation of a NS-NS merger remnant accretion disk,
we focus on understanding the evolution of the system in isolation to identify intrinsic features.
A more realistic system would include an initial distribution of dynamical ejecta around the
disk, which would modify the way in which both the jet and wind evolve (c.f., \citealt{fernandez_2017} for
a BH-NS system; see also \S\ref{s:initial_condition_uncertainties}}). 

\begin{table*}
\centering
\begin{minipage}{14cm}
\caption{List of evolved models and summary of results\label{t:models}\label{t:results}.
Columns from left to right show model name, dimensionality,
viscosity parameter, evolution time in seconds and in $r_g/c = GM_{\rm bh}/c^3\simeq 1.5\times 10^{-5}$\,s, 
unbound mass ejected at $10^9$\,cm
in units of the initial torus mass ($M_{\rm t0}=0.033M_\odot$ for GRMHD and $M_{\rm t0}=0.030M\odot$
for hydrodynamic models) and in $M_\odot$, 
average electron fraction and radial velocity of the outflow, unbound mass ejected that is 
lanthanide-poor ($Y_e>0.25$) in units of the initial torus and in solar masses, 
and average velocity of lanthanide-poor outflow.}
\begin{tabular}{lccccccccccccc}
\hline
{Model}&
{dim.} &
{$\alpha$} &
\multicolumn{2}{c}{$t_{\rm max}$\phantom{000}} & 
\multicolumn{2}{c}{$M_{\rm ej}\phantom{000}$} & 
{$\langle Y_e\rangle$} &
{$\langle v^r/c\rangle$} &
\multicolumn{2}{c}{$M_{\rm ej,LP}\phantom{000}$} & 
{$\langle v^r/c\rangle_{\rm LP}$}\\
{} & {} & {} & 
{(s)} & {($10^5\,r_g/c$)} & {($M_{\rm t0}$)} & {($10^{-3}M_\odot$)}& {} & {($10^{-2}$)} &
        {($M_{\rm t0}$)}  & {($10^{-3}M_\odot$)} & {($10^{-2}$)} \\
\hline
B3d     & 3D  & ...  & 9.3  & 6.3  & 0.39 & 13 & 0.16 & 11  & 0.04 & 1.2 & 22 \\      
\noalign{\smallskip}                                                  
h2d-v01 & 2D  & 0.01 & 26.5 & 18   & 0.16 & 4.8  & 0.26 & 2.7 & 0.06 & 1.9 & 3.1 \\      
h2d-v03 & 2D  & 0.03 & 16.5 & 11   & 0.21 & 6.3  & 0.20 & 3.2 & 0.04 & 1.1 & 4.5 \\      
h2d-v10 & 2D  & 0.10 &  8.8 & 5.9  & 0.22 & 6.7  & 0.17 & 5.2 & 0.03 & 0.8 & 10  \\      
\hline
\hline
\label{table:models}
\end{tabular}
\end{minipage}
\end{table*}

The initial equilibrium torus is constructed using a gravity consistent with
each method: full Kerr metric for the GRMHD model \citep{fishbone1976} or pseudo-Newtonian for
hydrodynamic models (e.g., F15). The torus has an initial rest mass $M_{t0}=0.033M_\odot$ in
the GRMHD model and $0.030M_\odot$ in the hydrodynamic models, 
constant specific angular momentum and constant entropy $s=8k_{\rm B}$ per baryon,
as well as constant initial electron fraction $Y_e = 0.1$.
The radius of the density peak is chosen to be $r_0 = 50\,\textrm{km}\simeq 11.3r_g$.
The ratio of internal energy to gravitational energy at the density peak is $15\%$,  
resulting in $H/R \simeq 0.35$ as generally obtained in full-physics simulations
of NS-NS mergers (e.g., \citealt{sekiguchi_2016}). For the equilibrium torus solution in GR,
this also corresponds to an inner radius of $0.62r_0$. The torus becomes optically-thin
to neutrinos within the first few orbits, justifying the approximations described in \S\ref{s:initial_model}.

A poloidal magnetic field is initially imposed in the GRMHD model
following \citet{sasha2011}. The
magnetic vector potential satisfies $A \propto r^{5/[3(\gamma_{\rm ad}-1)]^2}\rho^2$, 
and is set to a constant when it drops below $10^{-3}$ of its maximum value, to
prevent the magnetic field from reaching the low-density edge of the
disk \citep{sasha2011}. Then, the magnetic field is renormalized at
each point so it makes up a fixed fraction of matter plus radiation
pressure. Next, a magnetic vector potential is computed by integrating up
$B^r$ and used for re-computing an updated magnetic field.
This field is then normalized so that density-weighted ratio of matter and radiation pressure
to magnetic pressure in the disk is $\langle\beta_{\rm
  pl}\rangle=\langle P\rangle /\langle P_{\rm mag}\rangle \approx 100$, where
\begin{equation}
\label{eq:pmag_def}
P_{\rm mag} = \frac{1}{2}b_\mu b^\mu\, \left[\frac{M_{\rm bh} c^2}{r_g^3}\right].
\end{equation}
The advantage of recomputing the magnetic field as described above is
in obtaining a more uniform
distribution of $\beta_{\rm pl}$ in the disk; this is
helpful for resolving the MRI.
The resulting field configuration is shown in Figure~\ref{f:initial_conditions}. The maximum
field strength is approximately $4\times 10^{14}$\,G.
Given our spatial resolution, we resolve the most unstable wavelength of the poloidal MRI 
$\lambda_{\rm MRI}$ with at least 10 meridional cells ($r\Delta \theta$) over most of the 
equatorial plane, as also shown in Figure~\ref{f:initial_conditions}. Neutrino effects
are not expected to modify the usual ideal MHD stability criterion in the torus
\citep{foucart_2015,guilet_2017}.

The use of a finite volume method requires imposing a floor
of density and internal energy.
While a higher floor of density minimizes numerical problems
near the inner radial boundary close to the BH, it also interferes
with the launching of the wind if the mass in the outer computational
domain becomes comparable to the mass ejected. We therefore
adopt a floor of density that varies in both space and time.
The floor $\rho_{\rm f}$ initially follows a
power law with radius $\propto r^{-2}$, normalized so that
$\rho_{\rm f}= 10^{-5}\rho_{\rm max}$ at $r=r_g$ ($\rho_{\rm max}$
is the initial maximum torus density). As the torus evolves,
we decrease the density floor with time inside a transition radius 
$r_{\rm t} = 4r_0 \simeq 45r_g$ (following the approach of \citealt{Just+15}) 
to account for the fact that
the disk density decreases with time. The functional form adopted is
\begin{eqnarray}
\frac{\rho_{\rm f}(r,t)}{\rho_{\rm t}} =
\begin{dcases}
\left[1+\left(\frac{t}{t_{\rm v}} +1 \right)^{-2}
                \left(\left[\frac{r_{\rm t}}{r}\right]^2 -1 \right) \right] & \qquad r\le r_{\rm t}\\
\left( \frac{r_{\rm t}}{r}\right)^2 & \qquad r > r_{\rm t},
\end{dcases}
\end{eqnarray}
where $\rho_{\rm t}$ is the initial density floor at $r=r_{\rm t}$,
and $t_{\rm v}$ is a characteristic timescale over which the density at the ISCO decreases, 
approximately $40$ orbits at the density maximum for $\alpha = 0.03$. The
time exponent comes from an empirical fit to the time dependence of the
density at the inner boundary in F15. The floor of
internal energy density in the GRMHD model is set to 
\begin{equation}
\epsilon_{\rm f}(r,t) = 10^{-7}\rho_{\rm max}c^2\left[\frac{\rho_{\rm f}(r,t)}{\rho_f(r_g,t)}\right]^\gamma_{\rm ad}.
\end{equation}
Both floors are not allowed to decrease below $10^{-20}\rho_{\rm max}$ (density) 
and $10^{-20}\rho_{\rm max}c^2$ (internal energy density) in the GRMHD model. In the hydrodynamic models, 
a floor of specific internal energy $p_{\rm f}/\rho_{\rm f}$ is used, with $p_{\rm f}$ 
chosen to be about $10^{-14}$ of the value at the initial pressure maximum.

At $t=0$, the space surrounding the torus is filled with 
material with density $\simeq 1.5\rho_{\rm f}$, and a separate mass
fraction $X_{\rm atm}=1$, modeling an inert hydrogen
atmosphere ($Y_e=1$). This mass fraction is included in the 
NSE system of equations~(\ref{eq:mass_conservation_nse})-(\ref{eq:charge_conservation_nse}) for
continuity at the torus edges, but it is not available to form $\alpha$ particles (i.e., it
is subtracted from the right hand side of equation~\ref{eq:xalpha_fit}).

Neutrino and nuclear source terms are set to zero when $\rho < 10\rho_f$. Given
the difficulty of GRMHD schemes to recover primitive from conserved variables
when the magnetic field dominates the energy density by large factors (e.g., \citealt{gammie_2003}), 
we ignore in our analysis any regions for which
\begin{equation}
\frac{b_\mu b^\mu}{\rho c^2} > 100.
\end{equation}
Given that these highly magnetized regions (e.g., the center of the jet) are also associated with 
densities close to the floor value, we further ignore any part of the simulation for which $\rho < 10 \rho_f$.
{In practice, these restrictions affect primarily the part of the outflow with the highest kinetic
and electromagnetic energy, with very minor effects on the ejected mass distribution (\S\ref{s:relativistic_ejecta}).}

\begin{figure*}
\includegraphics*[width=\textwidth]{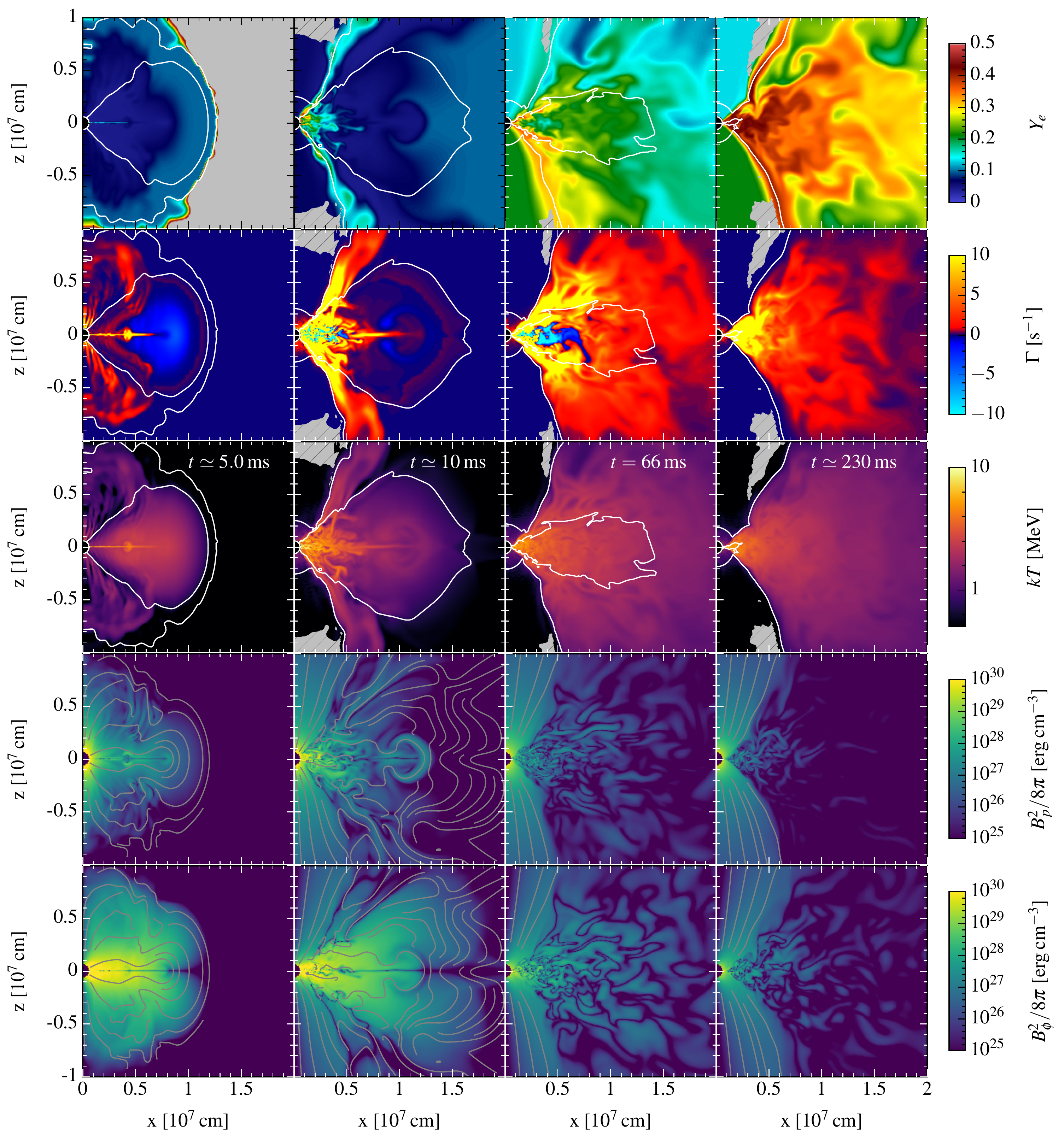}
\caption{Snapshots of the early evolution of the GRMHD model (slice $y=0$), with each column corresponding to
the time as labeled in the middle row (the orbital time at the initial density peak is $3.3$\,ms, or $224r_g/c$).
From top to bottom, rows correspond to electron fraction, neutrino number source term $\Gamma$ (equation~\ref{eq:Gnu}),
temperature, poloidal magnetic pressure, and toroidal magnetic pressure, respectively. The 
white contours correspond to mass densities of $10^6$\,g\,cm$^{-3}$  (outer) and $10^9$\,g\,cm$^{-3}$ (inner),
and some magnetic field lines are shown in gray in the lower two rows. The gray hatched area corresponds
to regions excluded from our analysis for having high magnetization or a density close to the 
floor value (\S\ref{s:initial_conditions}).}
\label{f:early_evolution_mhd}
\end{figure*}

%
\subsection{Models Evolved}
\label{s:models_evolved}

All production models are shown in Table~\ref{t:models}.  In addition to the GRMHD model B3d, 
we evolve three axisymmetric hydrodynamic models that differ only in the magnitude of
the $\alpha$ viscosity parameter: h2d-v01 ($\alpha=0.01$), h2d-v03 ($\alpha=0.03$), and
h2d-v10 ($\alpha=0.1$). Aside from the initial magnetic field and viscosity parameters,
all other properties in the initial tori are as close as possible between GRMHD and hydrodynamic
models given the different implementations of gravity (the GRMHD torus has an initial rest mass $\sim 10\%$ 
higher than the hydrodynamic models).

\section{Results}
\label{s:results}

\subsection{Early disk evolution}
\label{s:early_evolution}

\begin{figure*}
\includegraphics*[width=0.9\textwidth]{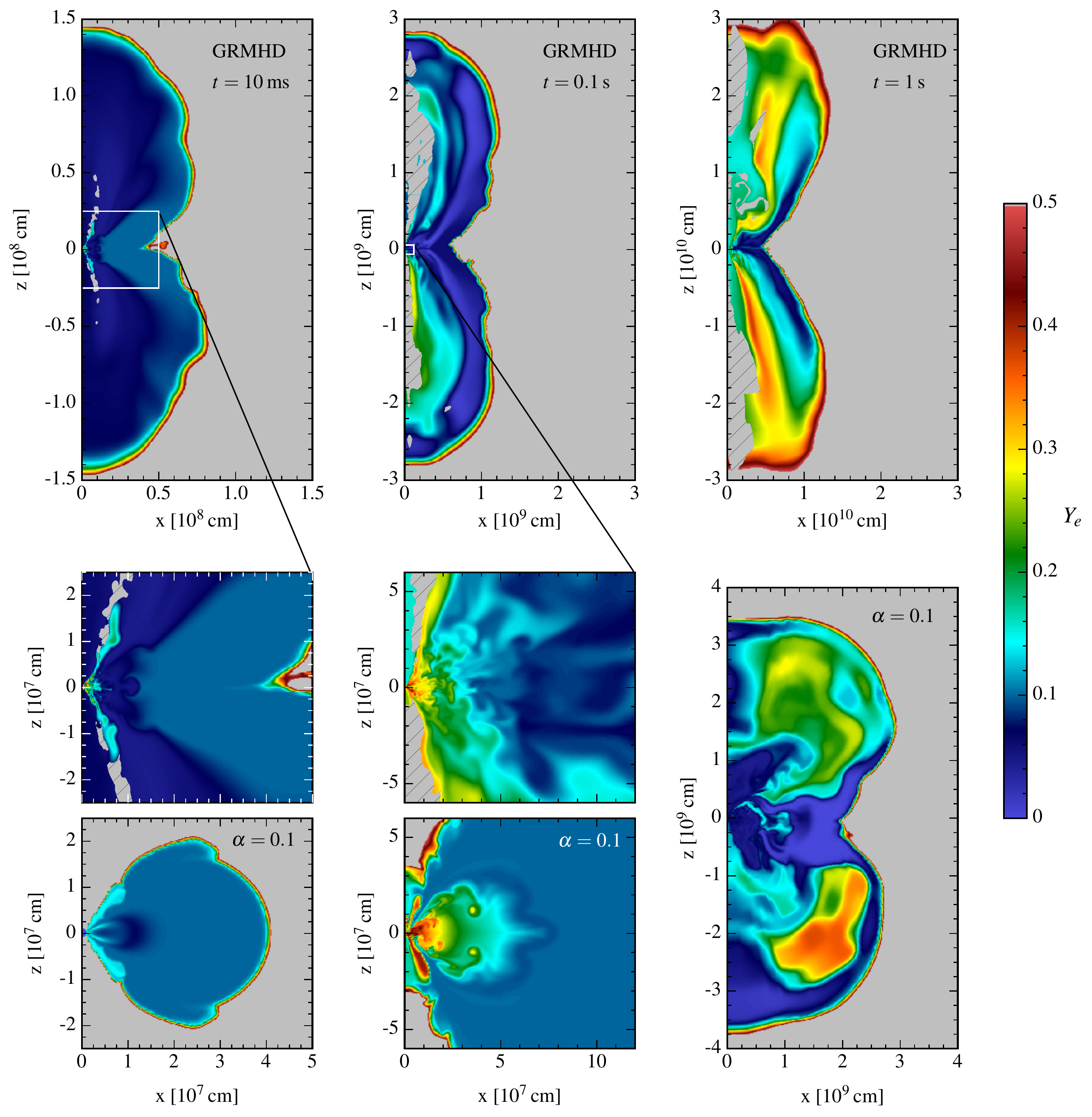}
\caption{Snapshots of the electron fraction in the GRMHD model (slice $y=0$, 
top and middle panels) and in the hydrodynamic model with $\alpha=0.1$ (bottom panels) 
at times $t=10$\,ms (left), $100$\,ms (center), and $1$\,s (right). 
The two leftmost panels in the middle row are zoom-ins of the corresponding panels in the top row.
Note the difference in scale between panels. The gray hatched region in the GRMHD model
is excluded from the analysis for having high magnetization or density close to the floor value (\S\ref{s:initial_conditions}).}
\label{f:ye_snapshots}
\end{figure*}

\begin{figure*}
\includegraphics*[width=\textwidth]{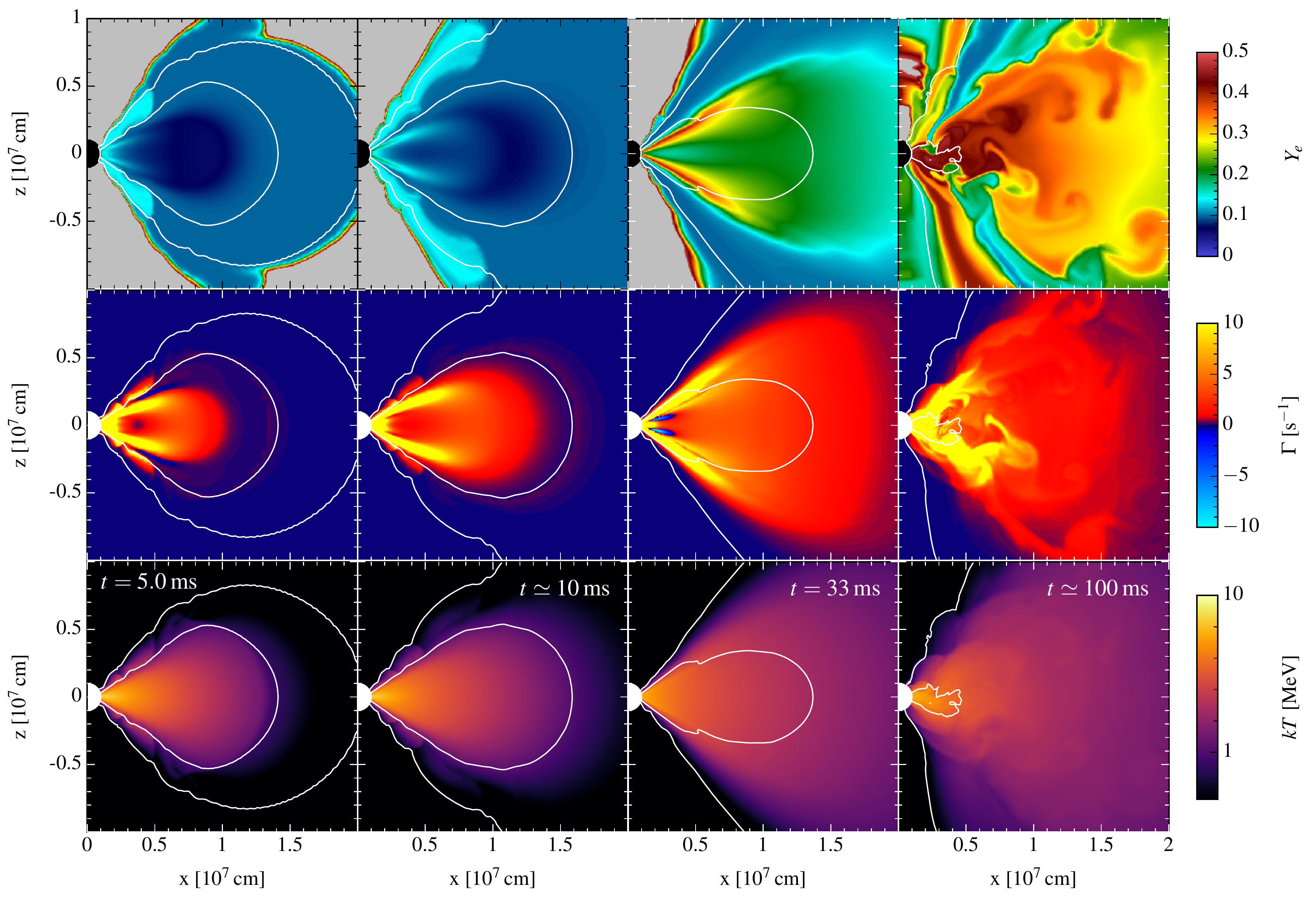}
\caption{Same as Figure~\ref{f:early_evolution_mhd}, but for the hydrodynamic model with $\alpha=0.1$.}
\label{f:early_evolution_hydro}
\end{figure*}

The initial dynamics of the GRMHD model (B3d) is determined primarily by
the choice of initial field geometry. Since the earliest stage of 
disk evolution leaves an imprint in the composition of the outflow,
we discuss its properties keeping in mind that if more realistic initial
conditions were used, it is unclear whether all of these features would be present.

The early evolution of the disk follows a sequence of stages
involving mostly laminar flow (e.g., \citealt{haw00}). When starting with a purely poloidal field,
orbital motion in the disk amplifies the toroidal magnetic field,
and the resulting radial gradients in magnetic pressure drive the
inner edge of the disk toward the black hole, starting the accretion process.
At the same time, the toroidal magnetic field changes sign on the disk equatorial plane,
generating a current sheet and a vertical magnetic pressure gradient 
that compresses material.

Figure~\ref{f:early_evolution_mhd} illustrates this early evolution,
which takes place within the first few orbits at the radius of the initial density peak 
(orbital period $t = 224\,r_g/c \simeq 3.3$\,ms). Turbulence driven by the MRI develops
by orbit 3 ($t\simeq 10$\,ms), starting the radial spread of the disk due to
sustained angular momentum transport.
Concurrent with the onset of accretion, a relativistic outflow is launched
around the rotation axis due the magnetic pressure gradient generated by field 
winding in the vicinity of the black hole (e.g., \citealt{hawley_2006}). In our model,
this expansion occurs on a few orbital times (Figure~\ref{f:early_evolution_mhd}).
Thereafter, the polar outflow is composed of an inner, magnetically dominated
jet around the rotation axis, and an outer matter-dominated outflow around it
(as described in, e.g., \citealt{devilliers_2005}).

The entropy in the disk is such that electrons are mildly degenerate,
resulting in an equilibrium electron fraction $Y_e\lesssim 0.1$ (equation~\ref{eq:ye_eq}),
not very different from the initial value. Before the onset of the MRI, neutrino cooling 
decreases the entropy and thus the equilibrium electron fraction except 
at the location of the current sheet, where dissipation results in heating
and an increase in  $Y_e$ (Figure~\ref{f:early_evolution_mhd}).
The onset of the vertical outflow removes low-$Y_e$ material from the neutronized
core of the disk and spreads it to large radii. Once the disk becomes
turbulent, the temperature increases in the inner regions,
and higher $Y_e$ material is launched along the
outer wall of the axial funnel. The spatial distribution of $Y_e$ in
this early outflow is shown in Figure~\ref{f:ye_snapshots}.

{\subsubsection{Comparison with hydrodynamic models}}

In contrast {to the GRMHD model}, hydrodynamic models display a mostly laminar evolution throughout
this time period, as shown in Figure~\ref{f:early_evolution_hydro} for the model
with $\alpha=0.1$ (h2d-v10). The absence of magnetic fields, the
event horizon, and 
general relativistic frame dragging precludes the
generation of a relativistic outflow in the polar direction. Angular momentum transport 
by the azimuthal shear stress causes the disk to begin accreting around orbit 1 
while simultaneously spreading the disk outward on its rear side. The GRMHD model 
does not begin this outward equatorial expansion until the MRI is fully 
developed (Figure~\ref{f:early_evolution_mhd}).
The overall geometry of the early outflow therefore differs significantly when
including magnetic fields, as illustrated in Figure~\ref{f:ye_snapshots}.
By $10$\,ms after the onset of the simulation, the hydrodynamic model remains
spheroidal, while the GRMHD has already expanded into a hourglass shape with
an approximately 3:1 ratio between vertical and horizontal dimensions.

The hydrodynamic model is subject to viscous heating from the beginning,
and therefore its temperature is higher than the GRMHD model during the first
few orbits, before the MRI-driven turbulence has time to develop. This 
results in an entropy difference between the two models, and therefore in
a different level of electron degeneracy and equilibrium electron fraction 
(equation~\ref{eq:ye_eq}). The hydrodynamic model gradually
raises its $Y_e$ throughout the disk, particularly in regions above
the equatorial plane and close to the inner edge of the disk, where the temperature
is high and the density is low. If the initial magnetic field at the 
time of torus formation is already turbulent (as found in, e.g., \citealt{kiuchi2014}), 
then the early thermal evolution of the disk might be closer to the hydrodynamic models
due to the enhanced energy deposition from turbulent dissipation (with implications
for the composition of the early outflow).

\vspace{0.25in}

\subsection{Long-term evolution, accretion history, and energy output}
\label{s:overview}

As the disk starts accreting onto the BH and spreading due to angular momentum 
transport, the average density decreases while the temperature remains
nearly constant {at a fixed position}, thereby gradually increasing the disk entropy, decreasing the
electron degeneracy, and thus increasing the equilibrium
value to which weak interactions drive $Y_e$. This change is visible in 
Figure~\ref{f:early_evolution_mhd}  through
the shrinking of the $10^9$\,g\,cm$^{-3}$ density contour and the increase
in $Y_e$ over $\sim 10-100$ orbits ($\sim 30-300$\,ms). The disk therefore starts
ejecting material with increasingly higher electron fraction, as visible
from Figure~\ref{f:ye_snapshots}. This process operates in both GRMHD and
hydrodynamic models, {as shown in Figure~\ref{f:early_evolution_hydro}}.

\begin{figure}
\includegraphics*[width=\columnwidth]{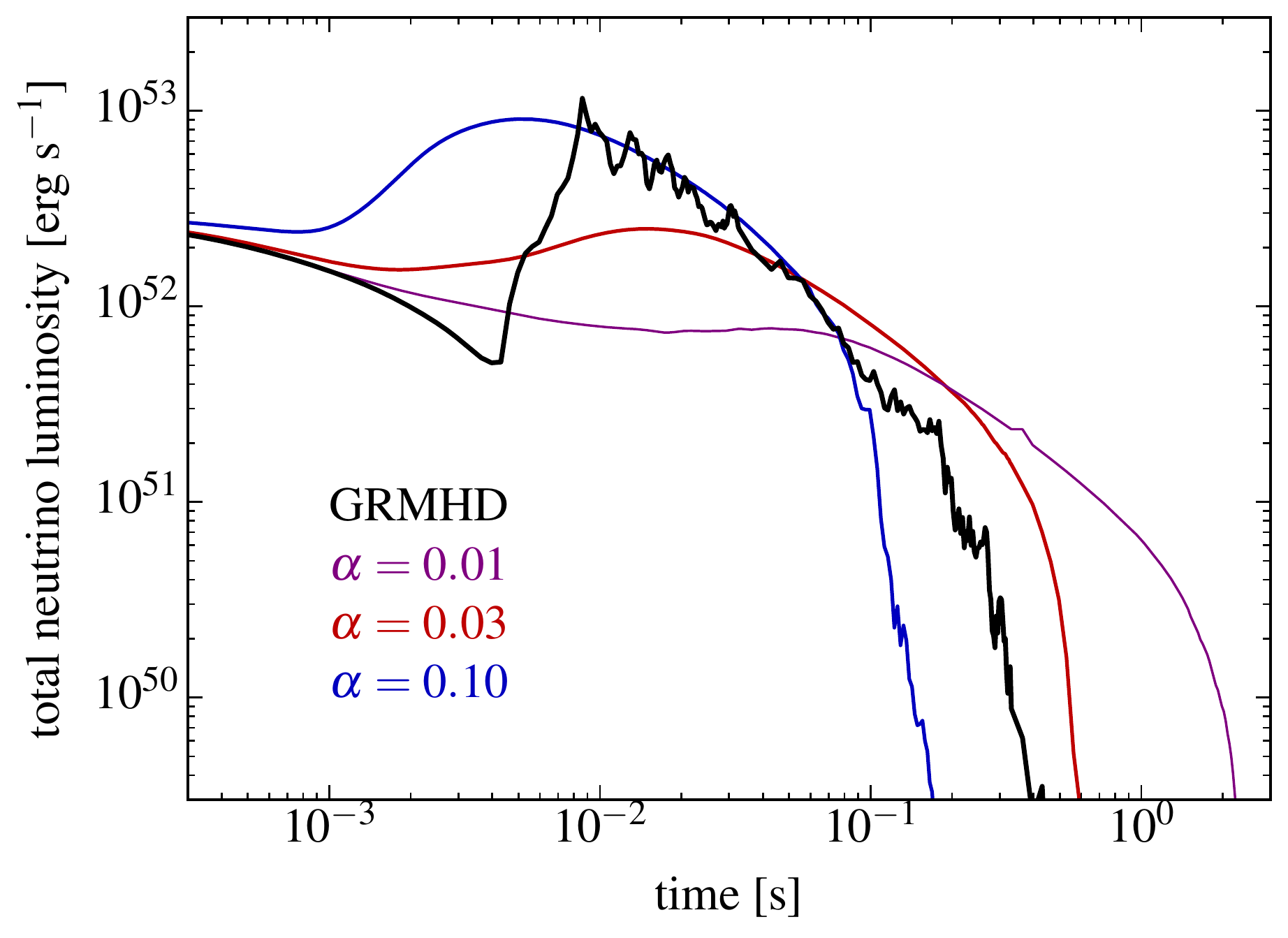}
\caption{Total neutrino and antineutrino luminosity (equation~\ref{eq:lnu_def}) as a function of time for the GRMHD model
and the three hydrodynamic models with different values of the
viscosity parameter, as labeled. For reference, each second of physical
time corresponds approximately to $6.7\times 10^4 r_g/c$.}
\label{f:lnu_hydro-mhd}
\end{figure}

The continued decrease in the density eventually causes weak
interactions to drop to dynamically
unimportant levels, thus freezing out $Y_e$. This transformation from a neutrino-cooled
disk \citep{popham1999,Chen&Beloborodov07} to an advection-dominated accretion 
flow \citep{Narayan&Yi94} occurs on the angular momentum transport timescale 
\citep{Beloborodov08,Metzger+09a}. This transition can be quantified by the evolution
of the total neutrino and antineutrino luminosity 
\begin{equation}
\label{eq:lnu_def}
L_\nu = \int \rho u_t Q\, \sqrt{-g}dx^r dx^\theta dx^\phi{,}
\end{equation}
{shown in} Figure~\ref{f:lnu_hydro-mhd}. 
The GRMHD model becomes advective over $\sim 300$\,ms, or about $100$ orbits.  

To further diagnose our models, we compute the rest mass flow rate at a given radius in the GRMHD model using
\begin{equation}
\label{eq:mdot_mhd_def}
\dot{M}(r) = \oint \rho u^r\,\sqrt{-g}dx^\theta dx^\phi{.}
\end{equation}
Similarly, the surface integral of the energy flux in the GRMHD model
is obtained from the stress-energy tensor
\begin{equation}
\dot{E}(r) = -\oint T^r_t\, \sqrt{-g}dx^\theta dx^\phi.
\end{equation}
Following \citet{mckinney2012}, we compute the kinetic,
enthalpy, and electromagnetic components separately:
\begin{eqnarray}
\label{eq:edot_ke_def}
\dot{E}_{\rm k}(r)    & = & -\oint \rho u^r (1+ u_t)\,\sqrt{-g}dx^\theta dx^\phi \\
\label{eq:edot_th_def}
\dot{E}_{\rm th}(r)    & = & -\oint (\epsilon + P) u^r u_t\,\sqrt{-g}dx^\theta dx^\phi \\
\label{eq:edot_em_def}
\dot{E}_{\rm EM}(r) & = & -\oint \left[ (b^\mu b_\mu)\, u^r u_t - b^r b_t\right] \, \sqrt{-g}dx^\theta dx^\phi,
\end{eqnarray}
respectively, where the notation follows the unit system of \S\ref{s:grmhd_methods}\footnote{The integrals are
carried out in code units and the result is multiplied by $M_{\rm bh}c^3/r_g$ to convert
to physical units.}. The rest mass power output at a given radius is $\dot{M}(r)\,c^2$. 

\begin{figure}
\includegraphics*[width=\columnwidth]{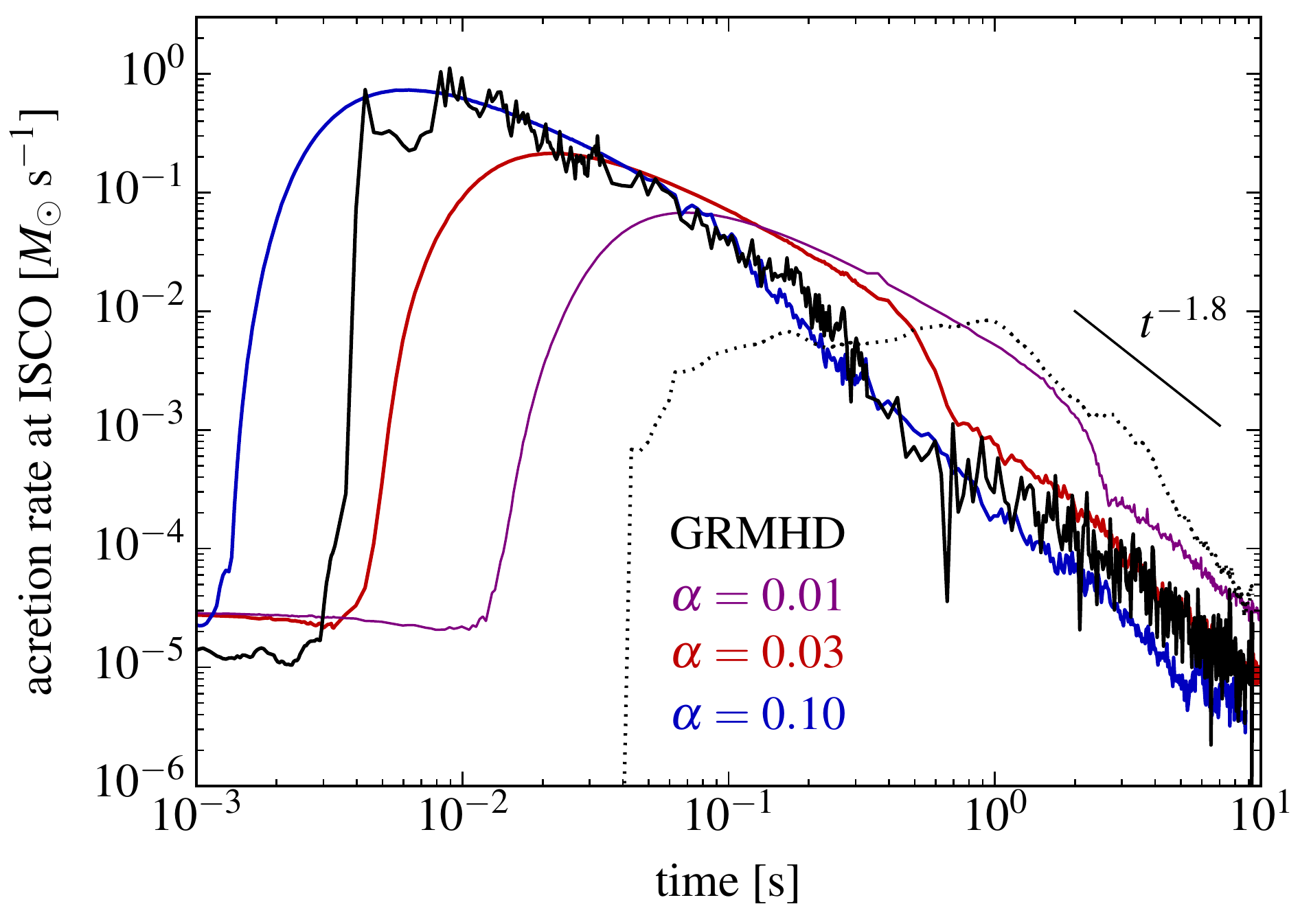}
\caption{Rest-mass accretion rate (equation~\ref{eq:mdot_mhd_def}) at the ISCO radius 
as a function of time for the GRMHD model and the hydrodynamic models with varying viscosity. 
The dotted black line shows the mass outflow rate at $10^9$\,cm for the GRMHD 
model (c.f. Figure~\ref{f:mout_hydro-mhd}). A power-law fit to the GRMHD accretion rate
gives $t^{-1.8}$ for $t>1$\,s, while the time dependence for both hydrodynamic models
with $\alpha=0.03-0.1$ is $t^{-1.9}$\,s.}
\label{f:mdot_hydro-mhd}
\end{figure}

\begin{figure}
\includegraphics*[width=\columnwidth]{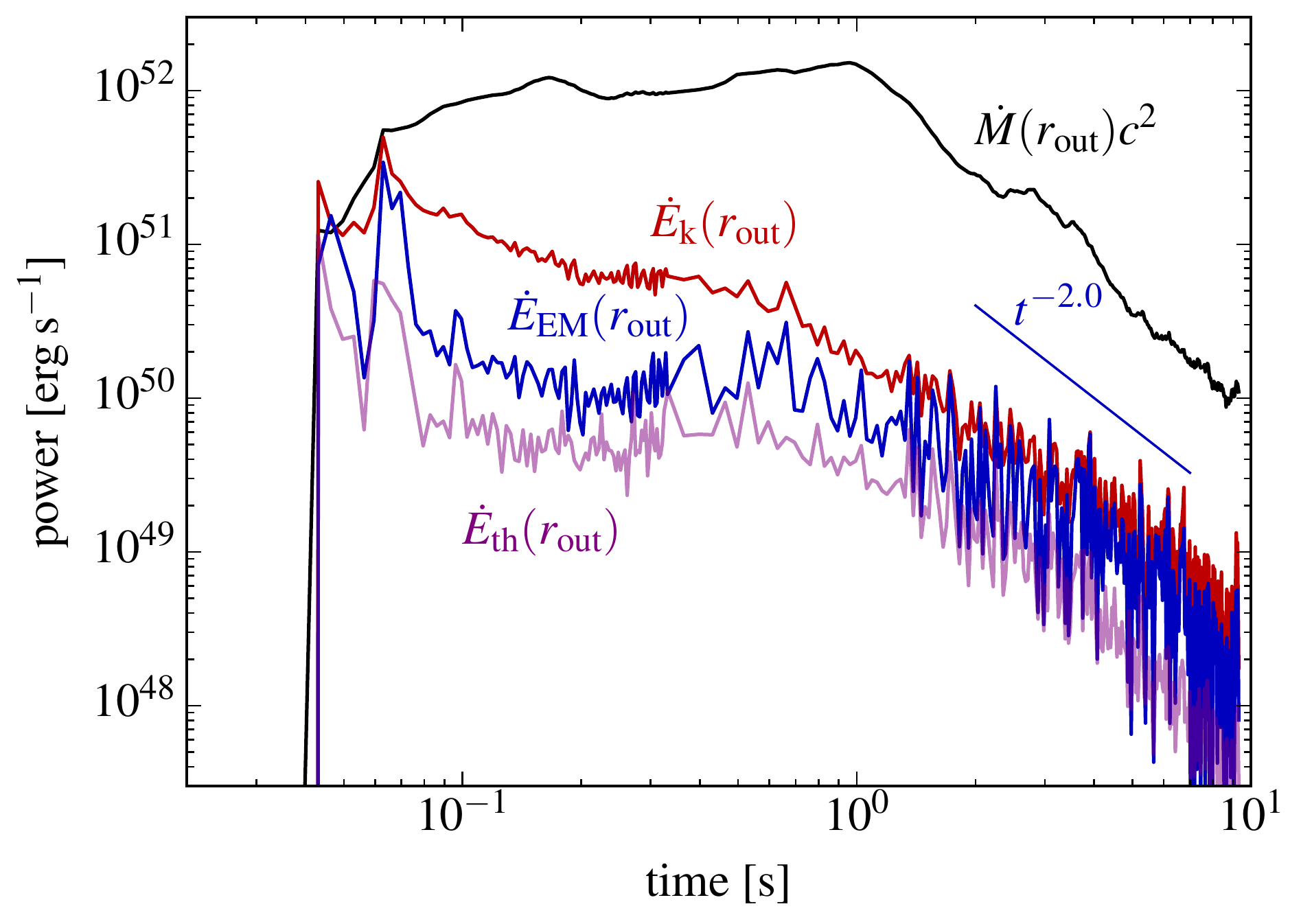}
\caption{Power generated by the GRMHD model at a radius $r_{\rm out} = 10^9$\,cm $\simeq 2,000r_g$, 
separated into components (rest mass, kinetic, electromagnetic, and thermal) 
according to equations~(\ref{eq:mdot_mhd_def}) 
and (\ref{eq:edot_ke_def})-(\ref{eq:edot_em_def}). The blue line shows
a power-law fit to the electromagnetic power for $t> 1$\,s.}
\label{f:edot_mhd}
\end{figure}

The rest mass accretion history at the ISCO radius is shown in Figure~\ref{f:mdot_hydro-mhd}. 
The mass flow into the BH reaches
a peak around the time MRI turbulence develops (c.f. Figure~\ref{f:early_evolution_mhd}), 
subsequently decreasing in intensity by almost five orders of magnitude by $t\simeq 10$\,s.
The total mass accreted in the GRMHD model is $0.02M_\odot$, or $\simeq 60\%$ of the initial torus mass.

The different components of the power  at a radius
$r_{\rm out} = 10^9$\,cm $\simeq 2,000r_g$ are shown in Figure~\ref{f:edot_mhd}.
The rest mass energy output dominates over all other components. The kinetic
power dominates over the electromagnetic power throughout the disk evolution.
We analyze more closely the mass ejection process and the properties of 
the ejecta in \S\S\ref{s:mass_ejection}-\ref{s:outflow_properties}.
The electromagnetic power output peaks at early times, when the magnetized
outflow generated during the early disk evolution 
(\S\ref{s:early_evolution} and Figure~\ref{f:early_evolution_mhd})
reaches $r=r_{\rm out}$. The total energy carried by the Poynting
jet (in all directions) is $\sim 3\times 10^{50}$\,erg over the duration of our 
simulation{\footnote{{The total energy generated by the jet depends on our choice
of high-magnetization and low-density cut (\S\ref{s:initial_conditions}). Removing
our cut increases the electromagnetic energy output to $1.3\times 10^{51}$\,erg.}}}. A power-law
fit to $\dot{E}_{\rm EM}(r_{\rm out})$ yields $t^{-2}$ for $t>1$\,s, roughly following
the time-dependence of the mass accretion rate. 

{\subsubsection{Comparison with hydrodynamic models}}

{The luminosity and mass flow rate for the hydrodynamic models is computed
using equations~(\ref{eq:lnu_def}) and (\ref{eq:mdot_mhd_def}) with $u_t = -1$, $\sqrt{-g} = r^2\sin\theta$,
and $(x^r, x^\theta,x^\phi)=(r,\theta,\phi)$.
Figure~\ref{f:lnu_hydro-mhd} shows that the early evolution of $L_\nu$ in the GRMHD model 
deviates somewhat from that of all the hydrodynamic models 
due to the delayed onset of angular momentum transport by the MRI relative to the viscous 
stress (\S\ref{s:early_evolution}). At late-times, the neutrino emission from the GRMHD model 
is bracketed by that of the hydrodynamic models with $\alpha=0.03-0.1$.} 

{As with the neutrino luminosity, the late-time accretion history of the GRMHD model is 
bracketed by the hydrodynamic models with $\alpha=0.03-0.1$. A power-law fit to the
accretion rate in the GRMHD model for $t>1$\,s yields $t^{-1.8}$, while in the
hydrodynamic models with $\alpha=0.03-0.1$ the dependence is $t^{-1.9}$. Despite the different 
treatment of 
gravity and processes driving angular momentum transport, the temporal slope of the mass 
accretion rate at late times is essentially the same in all models.}

{The main difference between the accretion histories of GRMHD and hydrodynamic models
has to do with the level of stochasticity of the fluid reaching the BH. Given that
MRI-driven turbulence transports angular momentum, mass flow onto the BH in the GRMHD model
shows fluctuations throughout its evolution. In contrast, accretion is smooth for the hydrodynamic
models for as long as neutrino cooling is important. Around the time when weak 
interactions freeze out, the magnitude of the accretion rate drops from its initial power-law 
evolution and becomes stochastic, latching onto a different power-law trajectory. While
the GRMHD models does not display such a marked transition in its accretion history,
fluctuations in the accretion rate show a visible modification around $t\sim 300$\,ms,
when neutrino cooling becomes unimportant.}

\vspace{0.25in}

\subsection{Mass ejection}
\label{s:mass_ejection}

\begin{figure}
\includegraphics*[width=\columnwidth]{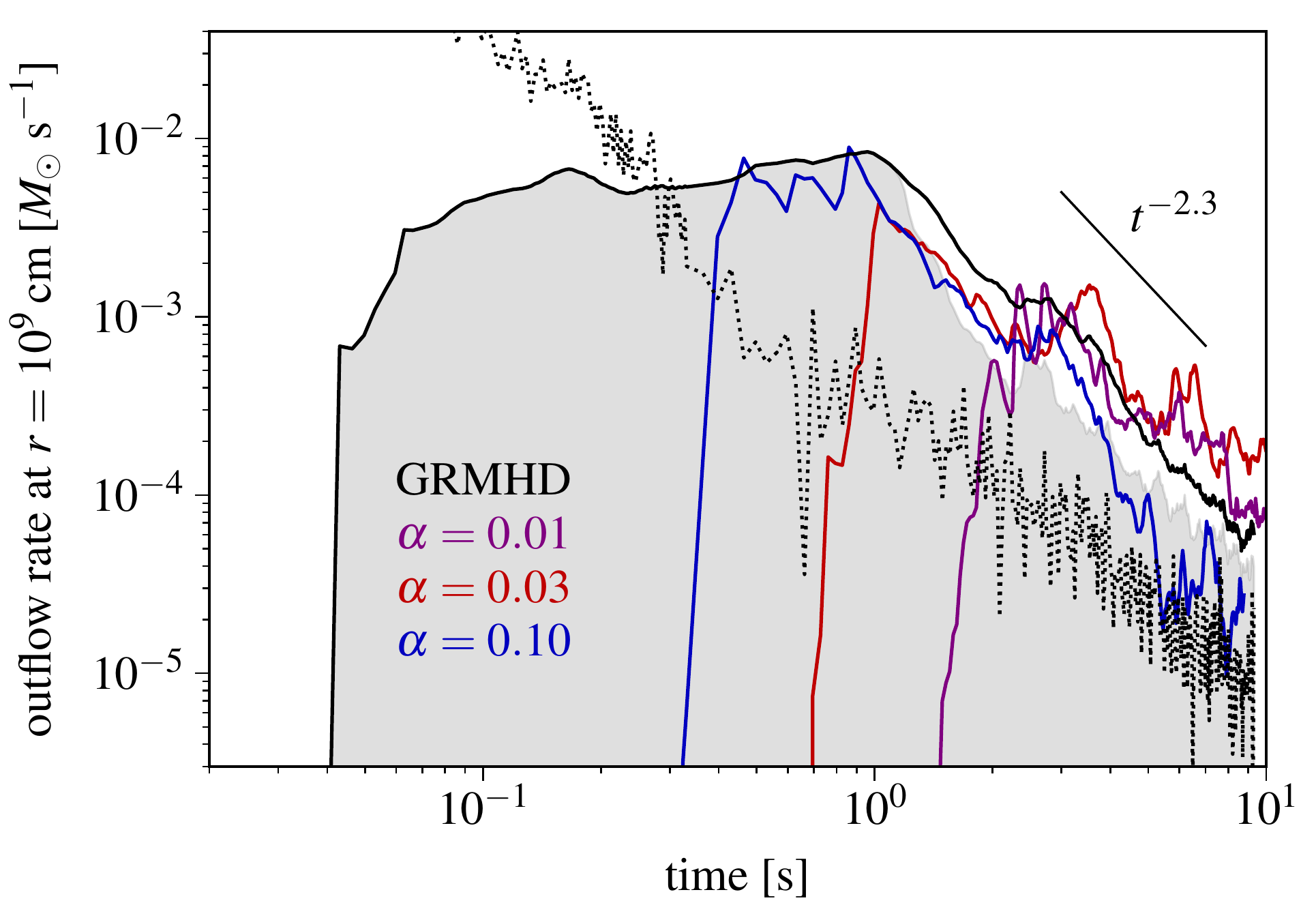}
\caption{Rest-mass unbound outflow rate (equation~\ref{eq:mdot_mhd_def} restricted to $-h\,u_t>1$) 
at $r_{\rm out}=10^9$\,cm  as a function of time for the GRMHD model (solid black line). 
A power-law fit to this mass-loss rate yields $t^{-2.3}$ for $t>1$\,s. For comparison, we also show
the outflow rates for the three hydrodynamic models with varying $\alpha$, as labeled.
The dotted line shows the mass accretion rate at the ISCO for the GRMHD model 
(c.f. Figure~\ref{f:mdot_hydro-mhd}), and the gray shaded area shows the fraction of 
the outflow in the GRMHD model that satisfies the condition $-u_t > 1$.}
\label{f:mout_hydro-mhd}
\end{figure}

The total amount of unbound mass ejected at a radius of $10^9$\,cm
is shown in Table~\ref{t:models}. In the GRMHD model, matter is considered to be
unbound when it satisfies the condition $-(1 + \gamma_{\rm ad} \epsilon/\rho)\,u_t = -h\,u_t>1$.
{This condition corresponds to a positive Bernoulli parameter in Newtonian gravity, 
accounting for
the internal energy available for conversion to kinetic energy via adiabatic expansion
upon subsequent evolution. For comparison, we also use the more restrictive `geodesic'
condition $-u_t > 1$, which corresponds to demanding that the 
escape speed be locally exceeded in Newtonian gravity, thus providing a lower limit on 
mass ejection (e.g., \citealt{kastaun_2015,bovard_2017}).}
The radius of $10^9$\,km ($\simeq 2,000\, r_g$) is chosen such that most of the outflow 
can be measured before the outer edge of the disk spreads to that point. 

The GRMHD model ejects about $1.3\times 10^{-2}M_\odot$, or $39\%$ of the
initial torus mass. The mass ejection history at $r=10^9$\,cm is shown
in Figure~\ref{f:mout_hydro-mhd}. The initial outflow reaches this radius by a time
of $\sim 40$\,ms, as can be seen from Figure~\ref{f:ye_snapshots}. This early
outflow plateaus at a time of $\sim 0.1$\,s, then slowly increases
to a peak at $t\sim 1$\,s. Thereafter, mass ejection decreases sharply
with time, following a $t^{-2.3}$ dependence. By $t=9.3$\,s the mass outflow rate 
is a factor $300$ lower than at its peak. At the end of the simulation, the rate of change of the 
cumulative ejected mass satisfies $d\ln M_{\rm ej}/d\ln t\simeq 0.03$, thus
\emph{mass ejection is {complete} to within other uncertainties}.

{Using the more restrictive `geodesic' criterion to determine the gravitational 
binding of the outflow results in only $30\%$ of the disk mass being ejected. 
Figure~\ref{f:mout_hydro-mhd} shows that nearly all the decrease (compared to the
Bernoulli criterion) arises in the late-time phase of the outflow, after $t=1$\,s.
At this time most of the material is ejected thermally by nuclear recombination
and dissipation of MHD turbulence given the absence of neutrino cooling 
(\S\ref{s:late_time_mass_ejection}). Since material is ejected from larger radii
in this phase, the outflow has not yet undergone full adiabatic expansion and
retains significant thermal energy at a radius $10^9$\,cm. The gravitational binding criterion
does not affect the total kinetic or electromagnetic energy of the outflow. Instead,
these quantities are dependent on the magnetization and low-density cut (\S\ref{s:initial_conditions}).}

\begin{figure*}
\includegraphics*[width=\textwidth]{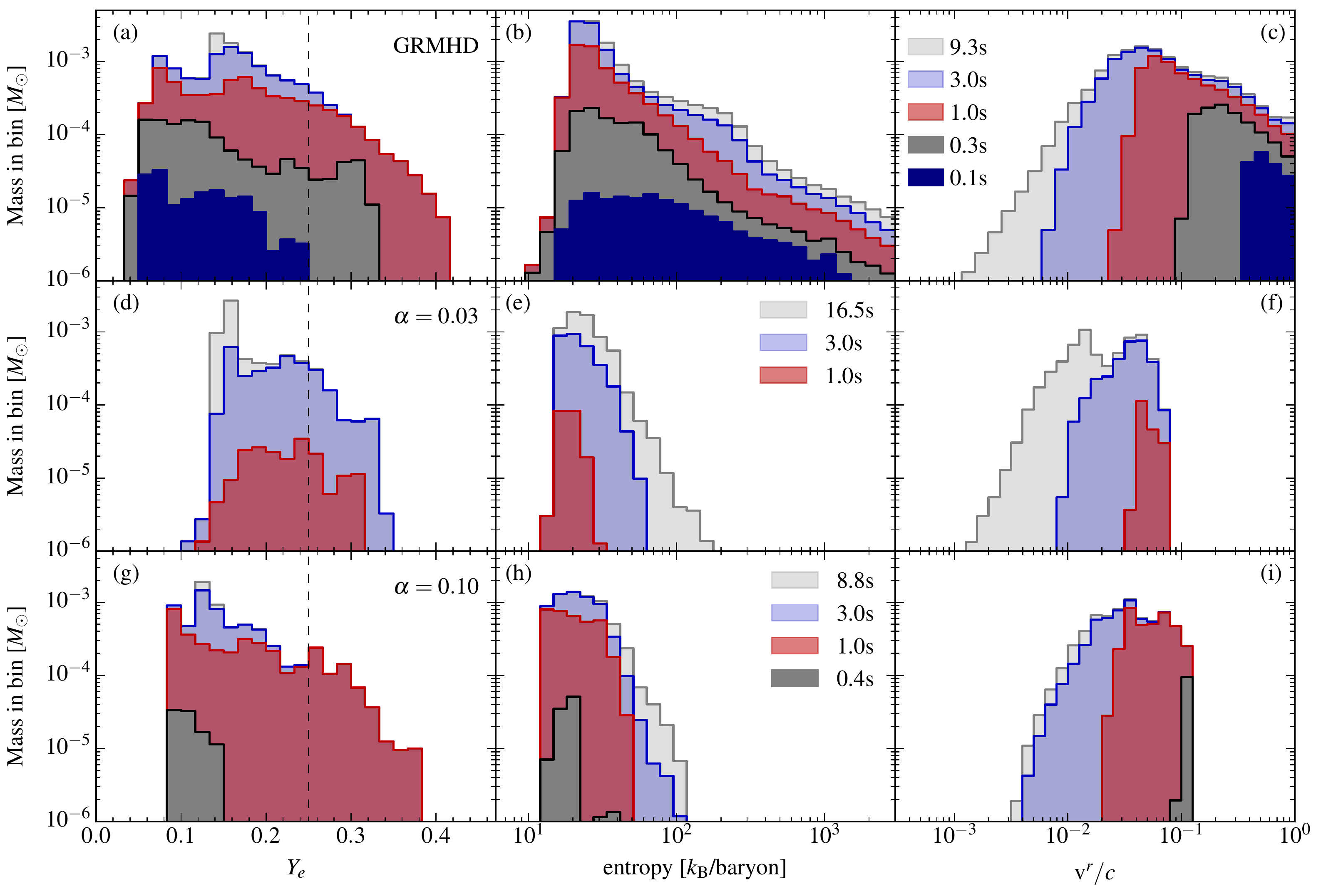}
\caption{Histograms of unbound ejected mass as a function of electron fraction 
(left column), entropy (middle column; equation~\ref{eq:entropy_definition}), 
and radial velocity $v^r$ (right column), 
measured at $r=10^9$\,cm, for the GRMHD model (top row), and the hydrodynamic models 
with $\alpha=0.03$ (middle row), and 
$\alpha=0.1$ (bottom row). Colors denote cumulative values at selected times, as labeled.
The vertical dashed line marks the approximate boundary between lanthanide-rich ($Y_e <0.25$)
and lanthanide-poor material (e.g., \citealt{kasen_2015}).
The bin sizes are $\Delta Y_e = 0.017$, $\Delta \ln s = 1.26$, and $\Delta \ln (v^r/c) = 1.31$.}
\label{f:mass_histograms}
\end{figure*}

The physics of the polar unbound outflows in three-dimensional GRMHD simulations
of accretion disks around spinning black holes has been studied by \citet{devilliers_2005} 
and \citet{hawley_2006}. They found that the jet core is magnetically-dominated,
with a very low matter density, and contains field lines that are primarily radial, with 
a degree of coiling that depends on the spin of the black hole. Matter outflow was
found to reside outside the jet `wall', being confined from the jet side by
centrifugal forces and on the outside by gas pressure from the disk corona.
When inspecting the effective gravitational potential, \citet{hawley_2006} found that
while centrifugal forces help direct the matter outflow, they do not accelerate it.
Instead, they argued that the combination of gas and magnetic pressure gradients in the corona
provide a suitable acceleration mechanism. Besides the narrow region around the jet wall, 
no unbound matter outflows were found, with bound coronal `backflows' closer to the
equatorial plane.

While our GRMHD model displays many of the features described in \citet{hawley_2006},
important differences arise. The accretion disk transitions from being neutrino cooled
at times $t\lesssim 100$\,ms ($\sim 10^4\,r_g/c$) to being advective, and
recombination of nucleons into alpha particles (equation~\ref{eq:qalpha}) provides an additional source
of energy for unbinding material at all angles. \citet{siegel_2018} have in fact argued
that nuclear recombination is the key process behind unbound mass ejection when including MHD.

Early mass ejection is confined to latitudes close to the rotation axis, and a configuration 
of an empty jet funnel plus matter wall is
indeed present. However, the matter outflow expands sideways significantly, wrapping
around behind the back of the disk (c.f.~Figure~\ref{f:ye_snapshots}). 
The disk itself spreads outward along the equatorial
plane due to angular momentum transport, and eventually also ejects mass from its
side opposite to the black hole. Most of the disk nevertheless remains gravitationally
bound, and at late times ($\gtrsim 1$\,s) some of this bound material makes its way
outwards through latitudes close to the equatorial plane. At the end of the simulation,
only $\lesssim 1\%$ of all the matter reaching a radius of $10^9$\,cm is gravitationally bound.

{\subsubsection{Comparison with hydrodynamic models}}

{For hydrodynamic models, we consider fluid with positive specific energy to be unbound.
These models eject $16-22\%$ of the initial disk mass in unbound material, with larger
ejecta for increasing viscosity parameter. The magnitude of this mass
ejection and its dependence on the viscosity parameter is consistent with 
previous hydrodynamic simulations of disks around spinning 
black holes (e.g., \citealt{Just+15}; F15). Figure~\ref{f:mout_hydro-mhd}
shows that the outflow starts much later than in the GRMHD model,
but exhibits a very similar temporal dependence once peak mass ejection
has been reached around a time of $1$\,s. As is the case with the neutrino
luminosity and accretion rate at the ISCO, 
the hydrodynamic models with $\alpha = 0.03$ and $\alpha=0.1$ bracket
the mass ejection history of the GRMHD model.}

{The difference in mass ejected and early temporal dependence between
GRMHD and hydrodynamic models arises from the absence of magnetically-driven
outflows at early times in the hydrodynamic models. In fact, the amount of
mass ejected before $1$\,s in the GRMHD model is about one half of the total:
$19\%$ of the initial torus mass before $1$\,s, and $20\%$ thereafter. The
latter is comparable to the mass ejected by the hydrodynamic models.}

{The primary mass ejection mechanism in the hydrodynamic models operates during
the advective stage, when viscous heating exceeds neutrino cooling (e.g., \citealt{Metzger+09a,Lee+09,FM13}). 
Angular momentum transport moves material to shallower regions of the gravitational
potential, from where energy injection by viscous heating and alpha particle recombination
can unbind it. This accounts for the late-time onset of mass ejection relative to
the GRMHD model. Mass ejection due to neutrino absorption (not included in our
simulations) is sub-dominant when a black hole is the central object (c.f. \citealt{Just+15}).}

\vspace{0.25in}

\subsection{Outflow properties}
\label{s:outflow_properties}

The distribution of electron fraction, entropy, and radial velocity in the
outflow at $r=10^9$ is shown in Figure~\ref{f:mass_histograms} for the GRMHD model.
Histograms are constructed by considering only unbound matter and excluding regions
with high magnetization and density close to the floor value (\S\ref{s:initial_conditions}).

The early phase of mass ejection in the GRMHD model is mostly neutron-rich ($Y_e \lesssim 0.2$)
and fast ($v^r/c > 0.1$). As time elapses, material with increasing electron fraction
and lower velocities enters the outflow up to about $1$\,s of evolution. At later times, the
trend toward lower velocities continues but the mean $Y_e$ decreases back to $0.1-0.2$. The
final electron fraction distribution spans the range $0.03-0.4$, and the outflow
velocity extends from $10^{-3}c$ up to relativistic motion. The average values
are $\langle Y_e\rangle = 0.16$ and $\langle v^r/c\rangle = 0.11$ (Table~\ref{t:models}).

The entropy of the outflow\footnote{The entropy is commonly used as one of the 
parameters that describes the $r$-process. Since the conditions during nucleosynthesis are
usually radiation-dominated, the entropy directly quantifies the number density of photons
and thus the strength of photodissociation, which is expected to balance neutron capture 
along the $r$-process path (Meng-Ru Wu \& St\'ephane Goriely, private communication).} is quantified by assuming 
that all matter species follow
an ideal gas distribution, consistent with our calculation of the temperature 
via equation~(\ref{eq:ptot_equation}):
\begin{eqnarray}
\label{eq:entropy_definition}
s & = & \left[\frac{5}{2} - \ln\left(\frac{n}{n_{\rm Q}} \right) \right]\left[1 + 
  Y_e - \frac{3}{4}X_\alpha\right] + \frac{4}{3}\frac{aT^3}{n k_{\rm B}} 
  -\left[Y_e\ln\left(Y_e\left[\frac{m_n}{m_e}\right]^{3/2} \right)\right.\nonumber\\ 
&& \left. + X_n\ln X_n + X_p\ln X_p +\frac{1}{4}X_\alpha\ln\left(\frac{X_\alpha}{32} \right)\right],
\end{eqnarray}
where $n = \rho/m_n$ is the baryon number density.
Figure~\ref{f:mass_histograms} shows that the entropy distribution of the GRMHD
simulation peaks in the range $20-30k_{\rm B}$ per baryon, and has an extended
tail to high values. This general shape is maintained throughout the evolution.

\begin{figure}
\centering
\includegraphics*[width=0.8\columnwidth]{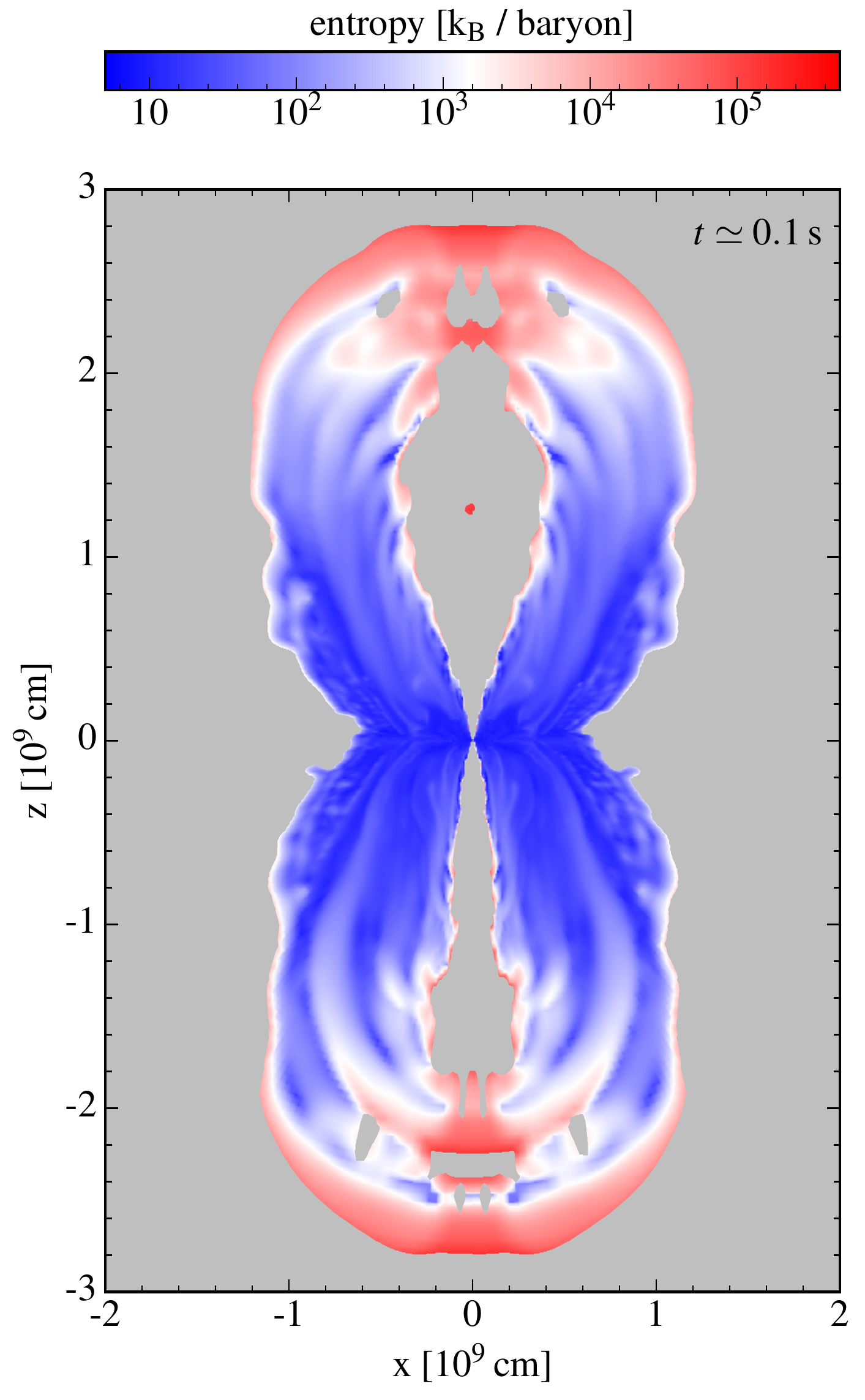}
\caption{Snapshot of the entropy (equation~\ref{eq:entropy_definition}) 
in the GRMHD model at time $t\simeq 0.1$\,s. The slice is perpendicular to the $y$-axis.
The gray-shaded region corresponds to regions excluded from our analysis for
having high magnetization or a density close to the floor value (\S\ref{s:initial_conditions}).}
\label{f:entropy_snapshot}
\end{figure}

The origin of the high-entropy tail becomes clear when inspecting Figure~\ref{f:entropy_snapshot}.
The bulk of the ejecta at mid-latitudes and on the equatorial plane has entropies below
$100k_{\rm B}$ per baryon. Much larger values are obtained around the jet head and funnel, 
and at the interface between the outflow and the ambient medium. The low densities
involved result in large entropies due to the radiation term $\propto T^3/\rho $ 
in equation~(\ref{eq:entropy_definition}), which dominates in this regime. Given that our numerical 
method has limited validity in highly magnetized regions close to the density floor, the results need 
to be interpreted with caution. For instance, the detailed form of the high-entropy tail
of the distribution is sensitive to our cut in magnetization and density (\S\ref{s:initial_conditions}).

{\subsubsection{Comparison with hydrodynamic models}}

{Figure~\ref{f:mass_histograms} also shows histograms for the hydrodynamic models that show
the closest similarity to the GRMHD model at late times. These models display a somewhat 
narrower distribution of electron fraction
relative to the GRMHD model, with a lower limit close to the initial value of $Y_e = 0.1$.
At the high-$Y_e$ end, increasing viscosity results in a tail extending to increasingly
higher $Y_e$, even thought the average of the distribution is lower (Table~\ref{t:models}).

The velocity distribution of the hydrodynamic models cuts off sharply around $v_r/c \sim 0.1$,
with a lower velocity tail that follows a power-law similar to the GRMHD model. The
cutoff value at high speed corresponds to the (Newtonian) escape speed at $\sim 500$\,km,
which is larger than the radius of the initial density peak by a factor $\sim 10$. A
larger radius of ejection is expected from the spreading of the disk in hydrodynamics
prior to the freezout of weak interactions. Also, the cutoff is consistent with the maximum 
velocity that can be obtained from the recombination of alpha particles, 
$\sqrt{2B_\alpha/m_\alpha}\simeq 0.12c$.}

{The low-entropy peak in the GRMHD model is similar to the entropy distribution of the hydrodynamic
models, and is thus expected if the ejection timescales and thermodynamic
properties of the outflow are similar (\S\ref{s:late_time_mass_ejection}).}

\vspace{0.25in}

\subsubsection{Relativistic Ejecta and Angular Distribution}
\label{s:relativistic_ejecta}

\begin{figure}
\centering
\includegraphics*[width=0.9\columnwidth]{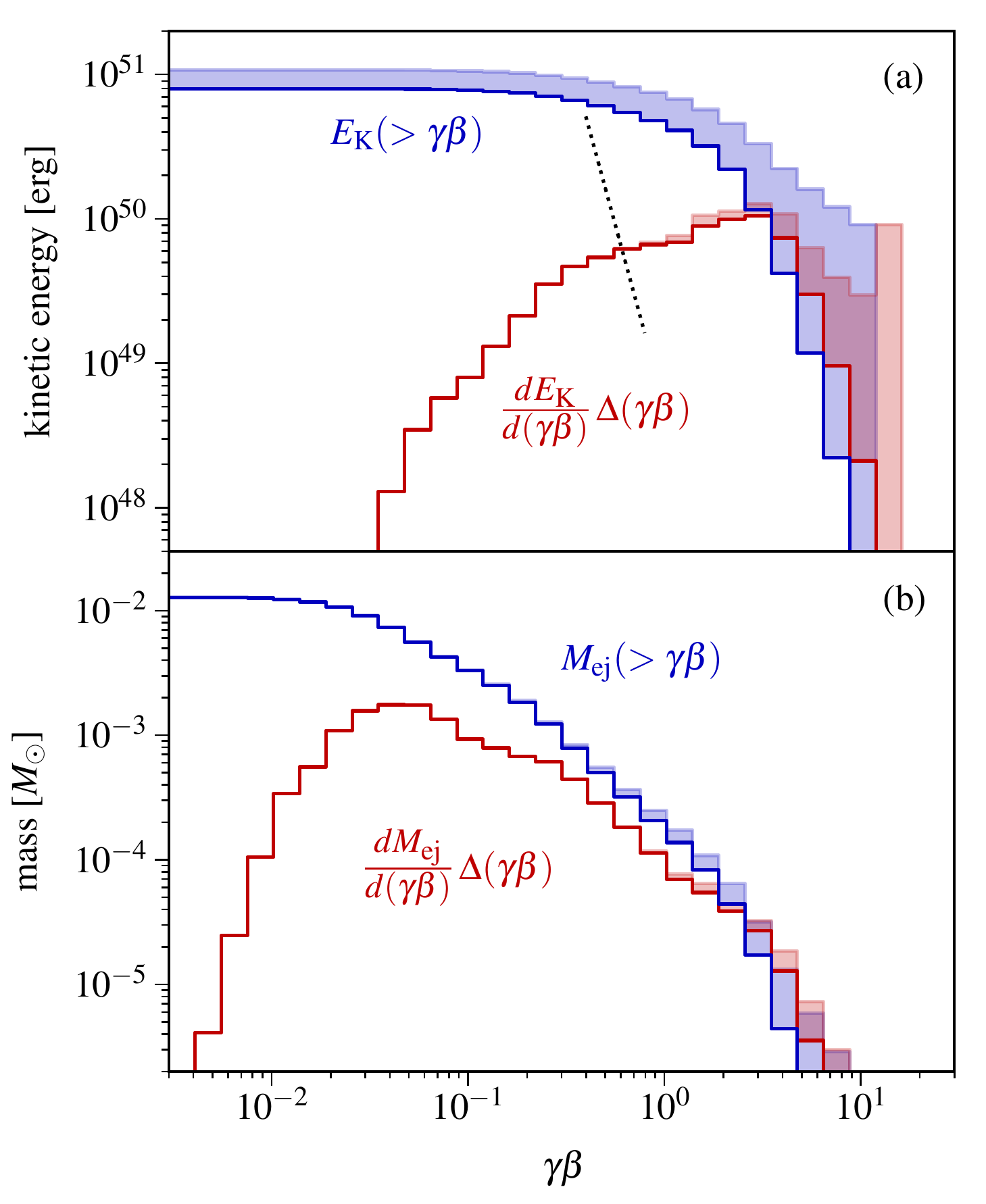}
\caption{Final histograms of kinetic energy (top) and unbound mass ejected (bottom)
as a function of relativistic momentum for the GRMHD model, as measured at $r=10^9$\,cm. 
In both cases, a quantity per bin and a reverse-cumulative version is shown 
(the bin size is $\Delta \ln\,\gamma\beta = 1.36$). The dotted line
is the spherical blast wave fit to the non-thermal emission from GW170817 by 
\citet{mooley_2018}, $E(>\gamma\beta) = 5\times 10^{50}\,(\gamma\beta/0.4)^{-5}$\,erg.
{The shaded areas indicate the results obtained when removing our low-density
and high-magnetization cut (\S\ref{s:initial_conditions}).}}
\label{f:histogram_gamma-beta}
\end{figure}

\begin{figure}
\centering
\includegraphics*[width=0.9\columnwidth]{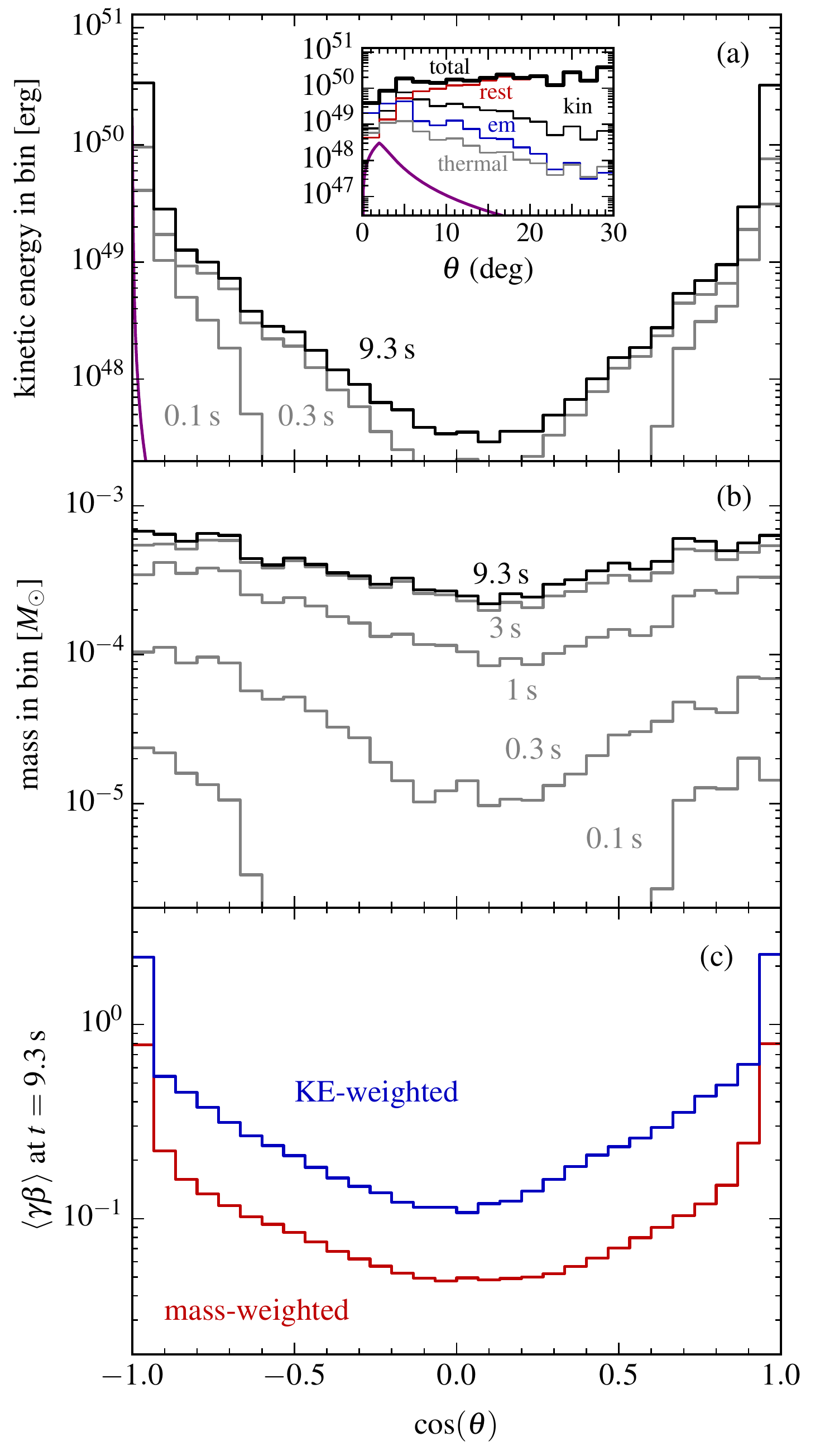}
\caption{Histograms of kinetic energy (top), unbound mass ejected (middle), and
average relativistic momentum (bottom) as a function of $\cos\theta$ for the GRMHD model,
as measured at $r=10^9$\,cm (the bin size is $\Delta\cos\theta = 0.067$; $21^\circ$ at the polar axis). 
Black and gray solid curves in the top two panels show cumulative values at different times, 
as labeled. The weights for the average momentum are the mass- and kinetic energy
histograms from panels (a) and (b). The purple line in the top panel shows the structured jet fit of 
\citet{davanzo_2018} to the non-thermal emission from GW170817, normalized to show energy per bin. 
The inset in the top panel shows energies per bin (rest mass: red, kinetic: thin black, 
electromagnetic: blue, thermal: grey, and total sum: thick black) as a function of polar angle 
close to the axis. {Note that the detailed angular distribution of the relativistic ejecta
will be sensitive to our choice of high-magnetization and low-density cut (\S\ref{s:initial_conditions}).}}
\label{f:histogram_angle}
\end{figure}

A small fraction of the ejecta from the GRMHD model achieves Lorentz
factors $\sim 1-10$. Figure~\ref{f:histogram_gamma-beta} shows
the final kinetic energy and mass histograms as a function of normalized relativistic
momentum\footnote{Not to be confused with the adiabiatic index $\gamma_{\rm ad}$
or the ratio of gas pressure to magnetic pressure $\beta_{\rm pl}$.} $\gamma\beta$.
While most of the mass has sub-relativistic velocity 
($\langle\gamma\beta\rangle\simeq 0.14$ weighted by mass), 
most of the kinetic energy of the outflow is carried by mildly relativistic material 
($\langle \gamma\beta\rangle\simeq 1.8$ weighted by kinetic energy).
As shown by Figure~\ref{f:edot_mhd}, the kinetic power exceeds the electromagnetic
power of the jet during most of the disk evolution. Note that the detailed
shape of the kinetic energy distribution for $\gamma\beta \gtrsim 3$ is set by
our choice of high magnetization and low density cut (\S\ref{s:initial_conditions}) when
constructing it{, as shown in Figure~\ref{f:histogram_gamma-beta}. The mass
distribution, on the other hand, is rather insensitive to our choice of cut}.

Figure~\ref{f:histogram_gamma-beta} also shows the reverse-cumulative distribution
of kinetic energy $E_k(>\gamma\beta)$ and of mass $M_{\rm ej}(>\gamma\beta)$. 
The former has been used as an input to models
that fit the non-thermal emission from GW170817 (e.g., \citealt{hotokezaka_2018b}). 
Comparing with the fit of \citet{mooley_2018}, our GRMHD model contains too much kinetic 
energy to account for the non-thermal emission when considered as a spherical blast wave.
Note however that the kinetic energy distribution is expected to change, as the
fastest component of the disk outflow will catch up with the dynamical ejecta
and interact with it.

The non-thermal emission from GW170817 can also be produced by a jet viewed
off-axis, which has an angle-dependent Lorentz factor 
(a `structured jet', e.g. \citealt{haggard_2017,lazzati_2018,margutti_2018}).
Figure~\ref{f:histogram_angle} shows histograms of mass, kinetic energy, 
and average relativistic momentum as a function of polar angle relative to the rotation axis.
The average is computed using both the mass and kinetic energy distributions
as weights. The former yields a lower value than the latter, as expected
from the momentum distributions shown in Figure~\ref{f:histogram_gamma-beta}.

Most of the kinetic energy of the outflow is concentrated in directions close
to the polar axis. The angular bins closest to the north and south directions 
($\sim 21^\circ$ in size) contain nearly all of the kinetic energy of the outflow,
approximately{\footnote{{Removing the high-magnetization and
low-density cut increases the total kinetic energy of the 
jet to $1.1\times 10^{51}$\,erg.}}} $8\times 10^{50}$\,erg. 
Moving away from the axis results in
a very steep decrease in the kinetic energy, with the equatorial direction
being lower than the poles by a factor of $\sim 1000$. This focusing of 
fast material along the polar direction is also reflected in the angular
distribution of relativistic momentum, both mass- and kinetic-energy-weighted.

In contrast, mass ejection shows a significant pole-equator anisotropy only
at early times $t\lesssim 0.3$\,s, with subsequent mass ejection turning
the distribution quasi-spherical, with a pole-to-equator anisotropy of
approximately $2:1$ by the end of the simulation.

For comparison, Figure~\ref{f:histogram_angle} also shows the structured
jet fit of \citet{davanzo_2018} for the isotropic equivalent energy of GW1710817, 
$E(\theta) = (\Delta \cos\theta/2)10^{52}/(1 + [\max(\theta,2^\circ)/2^\circ]^{3.5})$\,erg,
where the prefactor normalizes the isotropic equivalent energy to the
angular bin size $\Delta \cos\theta$. When considered over the entire range of polar angles,
the functional form of the fit has a much steeper decay with polar angle
than implied by our angular histogram. The inset of 
Figure~\ref{f:histogram_angle} shows a zoom-in on the angular distribution 
of all emitted forms of energy close to the axis. While the outflow
generated in our GRMHD model produces too much kinetic energy relative
to the non-thermal emission of GW170817, the angular dependence of the
kinetic, electromagnetic, and thermal components is compatible with the
fit of \citet{davanzo_2018}.
Again, we caution that
the fastest component of the disk outflow will almost certainly interact
with the dynamical ejecta and therefore the kinetic energy distribution
will change relative to that shown in Figure~\ref{f:histogram_angle} 
{(\S\ref{s:initial_condition_uncertainties})}

\subsubsection{Mass Ejection in the Advective Stage}
\label{s:late_time_mass_ejection}

\begin{figure*}
\includegraphics*[width=\textwidth]{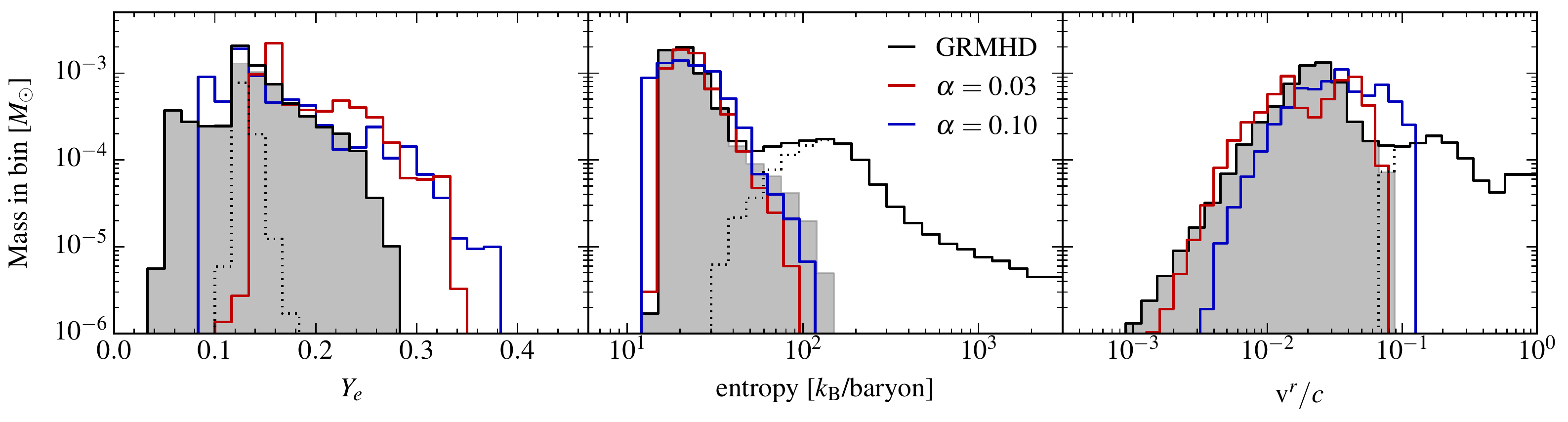}
\caption{Mass histograms of unbound matter ejected at $r=10^9$\,cm for $t>1$\,s by the 
GRMHD model B3d, as a function of electron fraction, entropy, and radial
velocity. The restriction to $t>1$\,s isolates outflows produced in the late-time advective phase.
For comparison, the total unbound matter ejected by hydrodynamic models 
with $\alpha=0.03$ and $0.1$ is also shown. The gray shaded area and dashed lines denote matter with $v^r/c < 0.1$ 
and $v^r/c\geq 0.1$ in the GRMHD model, respectively. The bin sizes are the same as in Figure~\ref{f:mass_histograms}}
\label{f:histogram_late}
\end{figure*}

Given that about half of all mass ejection by the GRMHD model
reaches our fiducial radius $r_{\rm out}$ after $1$\,s, and that 
the subsequent evolution of the 
mass ejection history (Figure~\ref{f:mout_hydro-mhd}) is similar
to that of the hydrodynamic models, it is worth exploring 
whether the properties of mass ejection are similar in
the GRMHD and hydrodynamic models once neutrino cooling has subsided.

Figure~\ref{f:histogram_late} shows histograms of mass
ejection after $1$\,s in the GRMHD model (obtained by subtracting
any prior contributions from the final histogram), together
with the total mass histograms for the hydrodynamic model.
Given that the amount of mass ejected after $1$\,s is comparable
to that in the hydrodynamic models, we expect the amplitudes
of the histograms to be similar.
The late-time electron fraction distribution has the same general
shape in all models, with the GRMHD model showing an overall
shift to low $Y_e$ (note that none of the models include neutrino 
absorption). 

The late-time velocity and entropy distributions in the GRMHD model
are bimodal, with the low-end distribution showing great 
similarity with the hydrodynamic models. Guided by the sharp cutoff
in the velocity distribution of the hydrodynamic models, we
also split the late GRMHD histograms into components with 
velocities lower and higher than $v^r/c = 0.1$. 

Figure~\ref{f:histogram_late} shows that the low-velocity component 
of the GRMHD model shows excellent agreement with the hydrodynamic 
models, pointing to an underlying similarity in the mass ejection mechanism.
The high-velocity component is also responsible for the high-entropy
tail of the GRMHD histogram at late times. Given its absence in the
hydrodynamic models, we surmise that it is associated with magnetic driving
close to the polar axis. This association is reinforced by
the $Y_e$ distribution of this fast component, which indicates
less reprocessing by neutrinos.

\subsection{Comparison with previous work}

The work of \citet{shibata2007} bears the most similarity to our
implementation of neutrino cooling and nuclear recombination (\S\ref{s:initial_model}). 
Since their 2D simulations were evolved for a relatively short amount of
time ($60$\,ms) given the decay of the MRI, we can only compare their
results with the earliest period in the evolution of our models.
Overall qualitative agreement is found in the growth time of the MRI and the
onset of accretion. Since their initial field strength is such
that $\beta_{\rm pl}=200$ and their initial tori are more massive 
($0.1-0.2M_\odot$) than ours, quantitative agreement in accretion
rates and neutrino luminosities is not expected.
Similar qualitative agreement in the early phase of MRI evolution
is found with the work of \citet{janiuk2013} and \citet{nouri_2017}

While the 2D neutrino radiation-MHD models of \citet{shibata2012}
cannot be directly compared with our coarser implementation
of neutrino physics, we can speculate about how inclusion of neutrino
absorption would affect our models. Their main result is that neutrinos are emitted
primarily along the polar funnel. Given that material in this 
region moves the fastest and therefore has a short expansion time,
the effect of neutrino absorption on the overall electron fraction
of the outflow might be further suppressed relative to that
in hydrodynamic models when a promptly-formed BH sits at the center and
the magnetic field is strong. If, on the other hand, the magnetic field
is initially weak, then the additional energy deposition in the polar
region can help energize a polar outflow (e.g. \citealt{just_2016,perego_2017}) 
and result in material with higher $Y_e$.

The work of \citet{siegel_2018} shows similarities and differences
with ours. While their equation of state, neutrino emission, and
nuclear recombination implementations are different, they find
the same fraction of accreted material as we do ($60\%$ of the initial disk mass). 
From this number, they extrapolate their $20\%$ of mass ejected within $400$\,ms
into an asymptotic fraction of $40\%$, which is consistent with our converged results.
Their disk also shows very similar accretion rates and
neutrino luminosity history as ours for the first $400$\,ms, which is expected given the
choice of disk mass and initial field strength, which is very similar
as well. While their final electron fraction distribution is somewhat 
narrower than ours at comparable evolutionary times, most of the outflow
in their model also leads to lanthanide-rich nucleosynthesis ($Y_e < 0.25$).

The main difference between our results and those of \citet{siegel_2018} 
is the significantly lower amount of ejected mass with $v^r/c > 0.25$ in their simulations
(Daniel Siegel, private communication). This can be attributed in part to 
our choice of initial field topology (\S\ref{s:initial_conditions}), which is optimal for
the generation of magnetically-dominated outflows \citep{sasha2011}. Also, while
their spatial resolution is comparable to ours at the initial torus location,
our grid configuration has a factor of at least $\sim 10$ higher resolution at the BH horizon and
can thus better capture the launching of relativistic material by magnetic fields.
Other factors such as the equation of state or neutrino treatment are less likely
to be important in accounting for this difference.

A larger body of literature exists on  magnetized merger simulations 
(e.g., \citealt{rezzolla_2011,kiuchi2014,paschalidis_2015,kiuchi2015,ruiz_2016}). 
Given the challenging nature of these calculations, important sacrifices
are normally made on the microphysics side, with implications for
the thermodynamics (i.e. non-inclusion of neutrino cooling). We therefore
refrain from comparing these results to ours.

{
\subsection{Uncertainties due to initial conditions}
\label{s:initial_condition_uncertainties}

Here we summarize the uncertainties associated with our specific choice of
initial condition. Our aim in this paper is to investigate the intrinsic properties of
merger remnant accretion disks when evolved in GRMHD over long timescales in a controlled setting.
We have therefore employed as initial condition an equilibrium torus surrounded by 
a low-density ambient medium (a `vacuum') and an idealized magnetic field geometry.

The most important choice we make is the initial field geometry. We have used a strong
poloidal field that generates an MRI that is easiest to resolve numerically, and which 
yields a robust jet. A more realistic field is likely to be not only dominantly toroidal
due to the merger dynamics, but also have a significant stochastic component that
might enhance turbulent dissipation from the beginning (e.g., \citealt{kiuchi2014}). This enhanced dissipation
can modify the thermal balance and result in different initial tori entropies and composition
relative to an unmagnetized merger. Purely toroidal magnetic fields have previously been found
problematic to generate jets (e.g., \citealt{devilliers_2005}), although it might just
be a matter of insufficient resolution \citep{liska_2018}. Jet-like structures have nevertheless been
observed to arise from mergers of neutron stars with interior poloidal fields \citep{rezzolla_2011,ruiz_2016}.

Using a more realistic distribution of dynamical ejecta, within which the torus is immersed,
is also expected to change the behavior of the outflow. The properties of the disk wind
can be modified due to mixing of fallback into the disk. The results of \citet{fernandez_2017} 
show that this admixture of neutron-rich fallback matter will result in a broader distribution
of electron fraction compared to evolving the disk alone. Because most of the dynamical ejecta
expands faster than the late-time thermal outflow, this interplay is not expected to affect 
the dynamics of this component of the wind. 

However, both the jet and fast component of the outflow, which can only obtained in GRMHD, 
have velocities similar or exceeding that of the dynamical ejecta, and will definitely
interact with it. The formation of a cocoon or even internal shocks are possible, with
consequences for the electromagnetic emission (\S\ref{s:obs_implications}). Such implications
must be addressed by further studies, in which the interaction between ejecta components
is the main focus.
}

\section{Observational Implications}
\label{s:obs_implications}

\subsection{Kilonova}

While the total ejecta mass in our GRMHD model ($0.013M_\odot$) is lower than that inferred
from the kilonova associated with GW170817 (e.g., \citealt{kasen_2017}), our initial disk 
mass is also lower than the value expected for GW170817. Depending on the equation of state and 
mass ratio, numerical relativity simulations predict a disk mass in the range $0.05-0.2M_\odot$ 
given the total inferred mass for GW170817 \citep{shibata_2017b}.

If we simply scale our ejecta to an initial disk mass of $0.1M_\odot$, we obtain 
a total ejecta mass of $0.04M_\odot$. Since most of the ejecta has $Y_e < 0.25$ 
with a mass-weighted velocity $v^r = 0.11c$ (Table~\ref{t:models}), the disk outflow can 
easily generate a red kilonova component similar to that from GW170817. 
On the other hand, the fraction of the total ejecta with $Y_e>0.25$ is
too small to account for the blue kilonova component. Performing the same
scaling to a disk of mass $0.1M_\odot$ would yield $3.6\times 10^{-3}M_\odot$ 
of lanthanide-poor ejecta, which is lower by almost an order of magnitude
relative to the required value. 

We caution, however, that these higher disk masses would be more opaque to
neutrinos, so it is probably not quantitatively accurate to simply scale
the results of our simulations to higher disk masses. Also, the mass
estimates from kilonova fits assume that the blue and red components
evolve independently of each other; proper radiation transport of 
the entire ejecta can yield different inferred ejecta masses \citep{kawaguchi_2018}.
Figure~\ref{f:ye_snapshots} shows that there is some degree
of spatial stratification in the electron fraction in addition to the
non-spherical distribution of ejecta, both of which can 
generate a viewing angle dependency for the kilonova.
On the positive side, the average radial velocity of
the lanthanide-poor material ($v^r = 0.22c$) is consistent with that from 
the blue kilonova from GW1710817. 

A small amount of lanthanide-poor mass ejected by a disk around a promptly-formed
BH is consistent with previous hydrodynamic disk models (\citealt{FM13,Just+15}; F15).
This has been used as an argument for a non-zero lifetime of the HMNS
in GW170817, which would irradiate the ejecta with neutrinos and increase
the electron fraction \citep{MF14,perego2014,fujibayashi_2018}, resulting
in a larger ejecta mass with $Y_e > 0.25$. Previous simulations of disks
around HMNSs find average outflow velocities that do not significantly
exceed $0.1c$, however, in conflict with the high velocity of the blue kilonova
from GW170817. A strongly-magnetized HMNS 
can produce {faster outflows (e.g., \citealt{Kiuchi_2012,Siegel_2014})}, 
resolving this tension \citep{metzger_2018}. Our results
show that this increase in outflow velocity is also realized
in GRMHD models of BH accretion disks, as the effects of magnetic stresses, turbulent energy dissipation,
nuclear recombination (and significant neutrino heating in the case of a HMNS) 
can all act in concert to accelerate matter.

\subsection{Neutron Precursor}

Mass with $v^r\sim c$ 
can generate strong early ($\sim$ hr timescales) thermal emission
powered by the radioactive decay of free neutrons
left over after the $r$-process completes \citep{metzger_2015}. To avoid
capturing neutrons, the ejecta must have an expansion
time shorter than about $5$\,ms \citep{goriely_2014,Just+15}. 

Given that our disk models achieve temperatures of $5\times 10^9$\,K (onset of nucleosynthesis) 
at radii $\sim 5\times 10^7$\,cm (e.g., \citealt{wu2016}), the required expansion times
are achieved for velocities $v^r\gtrsim 0.3c$. Using a conservative
lower limit of $0.4c$, we find that about $7.4\times 10^{-4}M_\odot$ of the
GRMHD disk outflow (about $2\%$ of the initial disk mass) achieves the required
velocities. Figures~\ref{f:mass_histograms} and \ref{f:histogram_late}
also show that the fastest ejecta has a significant neutron fraction
at both early and late times. This amount of mass is comparable to or
larger than the fraction of the dynamical ejecta found by \citet{bauswein2013}
to meet the conditions for a free neutron outflow. It is thus
possible that the disk outflow is critical for understanding the 
thermal emission from NS mergers on hour timescales. Note however that
a neutron-powered precursor from the disk outflow would require that the
material is not obscured by opaque dynamical ejecta in front of it, thus
realistic initial conditions are required to further assess the viability
of this counterpart.

\subsection{Non-thermal Emission}

As of this writing, the non-thermal emission from GW170817 still displays a single 
synchrotron power-law spectral energy 
distribution extending from the X-rays to the radio band, and a luminosity
that has reached a peak and is now decreasing \citep{alexander_2018,nynka_2018}. 
Models that can account for this behavior
include an off-axis jet with an angle-dependent Lorentz factor (a 
`structured jet') or a quasi-spherical blast wave with radial 
structure.

The fastest ejecta in our GRMHD model has
too much kinetic energy relative to models that can fit the 
observations of GW170817 (\S\ref{s:relativistic_ejecta}). Our initial magnetic field is
poloidal in shape and strong in magnitude ($\sim 4\times 10^{14}$\,G),
which optimizes the conditions for the emergence of a relativistic outflow.
Use of other initial conditions, including a weaker field with
a significant toroidal component, will likely result in a weaker
jet and lower kinetic power output. 

\section{Summary and Outlook}
\label{s:summary}

We have performed long-term, 3D GRMHD simulations of BH
accretion disks formed during neutron star mergers. Our models start with
an equilibrium torus, a strong poloidal field, and make 
use of a suitably calibrated gamma-law equation of state, with 
approximations for the temperature, neutrino cooling, and nuclear
recombination that account for the dominant effects of realistic microphysics
on the dynamics and composition
of the flow. These approximations enable us to maximize the amount of 
physical time simulated and therefore achieve {completion of} mass
{ejection} to large radii.
To connect with previous work {and to better diagnose the GRMHD results}, we have also carried out 2D hydrodynamic 
simulations with shear viscosity and a pseudo-Newtonian potential, and identical
treatment of other physics. Our main results are the following:
\newline

\noindent 1. -- When including MHD effects in general relativity, the total mass
	        ejected from the disk is $40\%$ of the initial torus mass 
                ($0.013M_\odot$ for an initial torus mass of $0.033M_\odot$; Table~\ref{t:models}
	        and Figure~\ref{f:mout_hydro-mhd}). This is larger by a factor of two relative to 
		hydrodynamic models. 
                \newline

\noindent 2. -- The ejected mass in the GRMHD model displays a broad distribution of electron fraction,
	        entropy, and radial velocity (Figure~\ref{f:mass_histograms}). The
	        majority of the outflow has $Y_e < 0.25$ and will thus result in
	        a lanthanide-rich composition. When scaling our ejected fractions to 
	        a disk mass of $0.1M_\odot$, the outflow from the GRMHD model can easily
	        account for the (nominal) mass and velocity of the red kilonova from GW170817.
	        The disk does not eject sufficient material with $Y_e>0.25$ to account
	        for the blue kilonova, however, despite achieving the right velocity
	        ($0.22c$, Table~\ref{t:models}). We caution, however, that our treatment
	        of weak interactions is approximate and does not include neutrino absorption.
		Also, the composition of the early outflow will depend more sensitively
	        on the initial distribution of $Y_e$ produced during the merger, which may be
		on average higher than the uniform and low value we assume ($0.1$). Finally,
	        the appearance of the kilonova can depend on the details of the spatial distribution
	        of the lanthanide mass fraction (e.g., \citealt{kasen_2017}), which in turn depends
		on the degree of mixing and/or stratification of the electron fraction, not
		just on the bulk amounts above or below $Y_e \sim 0.25$.
                \newline

\noindent 3. -- Mass ejection in MHD can be divided into two phases: an early ($t\leq 1$\,s as
	        measured from $r=10^9$\,cm), 
		magnetically-mediated phase, absent in hydrodynamic models, and a late phase 
	        following freezout of weak interactions (Figures~\ref{f:early_evolution_mhd} 
	        and \ref{f:lnu_hydro-mhd}), which operates on the 
                angular-momentum transport timescale.
                The slow component ($v^r/c<0.1$) of the late-time outflow
	        shows similar properties in both MHD
	        and in hydronamic models (Figure~\ref{f:histogram_late}). This similarity points to a shared 
		mechanism for mass ejection: neutrino cooling freezes out and thermal 
                energy is deposited by turbulent dissipation or viscous heating, and
	        by recombination of nucleons into alpha particles. The GRMHD model has an
		additional fast component at late times which must be mediated by
	        magnetic processes given its absence in hydrodynamic models.
		The accretion history, late-time mass ejection, and neutrino luminosity of 
	        the GRMHD model is bracketed by the hydrodynamic models that use 
		$\alpha=0.03$ and $\alpha=0.1$ (Figures~\ref{f:lnu_hydro-mhd}, \ref{f:mdot_hydro-mhd},
		and \ref{f:mout_hydro-mhd}).
                \newline

\noindent 4. -- Given our initial field geometry, which is optimized for efficient extraction
		of energy from the BH, we obtain a robust jet carrying
	        $3\times 10^{50}$\,erg of electromagnetic energy (in all directions; Figure~\ref{f:edot_mhd}). A small fraction
	        of the ejecta achieves relativistic velocities at latitudes close to the
	        rotation axis (Figure~\ref{f:histogram_gamma-beta} and \ref{f:histogram_angle}).
	        Comparing with models that fit the non-thermal emission from GW170817 shows
	        that our model contains too much kinetic energy in matter with
		$\gamma\beta\gtrsim 1$ (\S\ref{s:relativistic_ejecta}). The angular dependence
	        of our jet is compatible with off-axis fits to the non-thermal emission from
	        GW170817 (Figure~\ref{f:histogram_angle}){.}
	        We caution that this component of the disk outflow will almost certainly
	        interact with the dynamical ejecta, and thus the final kinetic energy
	        distribution in a realistic setting will likely differ from that in our models.
		\newline

\noindent 5. -- A few percent of the ejecta in the GRMHD model ($7\times 10^{-4}M_\odot$) 
		also has sufficient velocity and low $Y_e$ to generate free neutrons not
		captured onto nuclei during the $r$-process
	        (\S\ref{s:obs_implications}). This component can generate early ($\sim $ hr
		timescale) thermal emission preceding the kilonova. The existence of this component
	        is likely to also be sensitive to the initial magnetic field configuration,
	        and on its location relative to the dynamical ejecta.
		\newline

Our GRMHD model can be improved in many ways to achieve a more realistic result. The
simplest modification is changing the initial field geometry, which is bound to have the 
largest impact on the fastest portion of the ejecta. We have adopted a strong poloidal
field with a topology that maximizes energy extraction from the BH. A more realistic
disk is expected to contain a significant toroidal component (e.g., \citealt{etienne2012,kiuchi2014}) 
that can be highly turbulent from the time of disk formation (and thus generate
a different thermal evolution than our equilibrium initial conditions).

A more challenging improvement involves a realistic treatment of neutrinos (emission
and absorption) and microphysics (EOS with full nuclear recombination). For the case
of a promptly-formed BH, the effect of neutrino absorption on the bulk of the outflow 
dynamics is likely to be secondary, given that (1) in hydrodynamic models, for which the 
outflow is slow and similar to the slow MHD component at late times, absorption is unimportant,
and (2) inclusion of magnetic fields adds a fast component which neutrinos have
a smaller chance to modify given its rapid expansion. Nevertheless, neutrino absorption is crucial
for determining a reliable composition and therefore detailed $r$-process yields
for comparison with observations (as demonstrated by \citealt{siegel_2018} who assess the
effect of neutrino absorption in post-processing). Also, in the case of weak magnetic fields, neutrino
energy deposition can combine with magnetic stresses in launching a successful 
jet (e.g., \citealt{just_2016,perego_2017}), and the magnitude of this energy deposition is 
very sensitive to the neutrino scheme adopted \citep{foucart_2018}.
If a HMNS survives for longer than a dynamical time, the effect of neutrinos become dynamically important 
and proper neutrino transport is essential  for a complete description.

Finally, more realistic initial conditions for density, temperature, and electron fraction
can be obtained by mapping from a dynamical merger simulation (as in \citealt{nouri_2017}). 
While the details of the initial thermodynamics and composition have a relatively 
minor ($\sim 10\%$) effect on the late-time slow disk outflow composition 
\citep{fernandez_2017}, the prompt MHD outflow preserves the initial composition of the disk.
Also, the mapping from binary parameters (total mass, mass ratio, 
spins, etc.) to initial disk mass and field geometry is non-trivial and essential 
for detailed comparison with observations. Finally, evaluating the feasibility of
models that rely on the interaction of a jet with the dynamical ejecta 
(such as models involving cocoon emission; e.g. \citealt{nagakura_2014,lazzati_2017,gottlieb_2018}) requires
proper initial conditions for all merger remnant components.

\section*{Acknowledgments}

We thank Meng-Ru Wu, Yong-Zhong Qian, and St\'ephane Goriely for helpful discussions.
We also thank Daniel Siegel for providing additional information about published simulation results,
and Austin Harris for help with running simulations.
{The anonymous referee provided constructive comments that improved the presentation of the paper.}
RF acknowledges support from the Natural Sciences and Engineering Research Council (NSERC)
of Canada, and from the Faculty of Science at the University of Alberta.
AT acknowledges support from Northwestern University.
EQ was supported in part by a Simons Investigator
award from the Simons Foundation, and the David and Lucile Packard Foundation.  
This work was also supported in part by the Gordon and Betty Moore Foundation through 
Grant GBMF5076.
Support for this work was provided by NASA through Einstein Postdoctoral
Fellowship grant numbered PF4-150122 (FF) and PF3-140115 (AT) awarded by the Chandra X-ray Center,
which is operated by the Smithsonian Astrophysical Observatory for NASA under
contract NAS8-03060, and through grant 80NSSC18K0565 (FF, AT).
DK is supported by the U.S. Department of Energy,  Office of Science, Office of Nuclear Physics, under contract numbers
DE-AC02-05CH11231, DE-SC0017616, and DE-SC0018297.
The software used in this work was in part developed by the DOE NNSA-ASC OASCR Flash Center at the
University of Chicago.
This research used resources of the National Energy Research Scientific Computing
Center (NERSC), which is supported by the Office of Science of the U.S. Department of Energy
under Contract No. DE-AC02-05CH11231. Computations were performed at
\emph{Carver} and \emph{Edison} (repositories m1186, m2058, m2401, and the \emph{scavenger} queue).

\appendix

\section{Microphysics Approximations}
\label{s:microphysics}

\subsection{Adiabatic Index}
\label{s:adiabatic_index}

\begin{figure*}
\includegraphics*[width=0.49\textwidth]{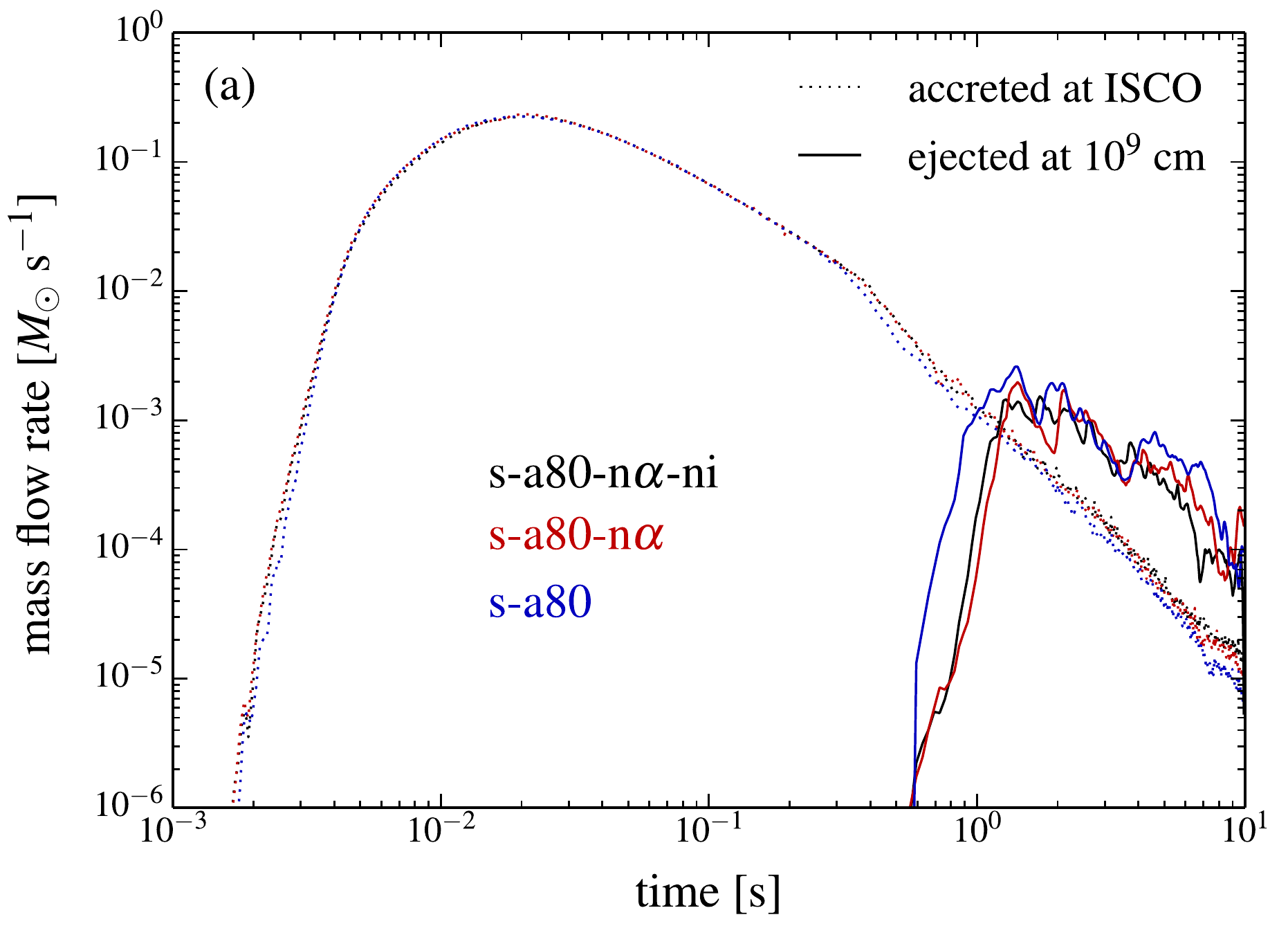}
\includegraphics*[width=0.49\textwidth]{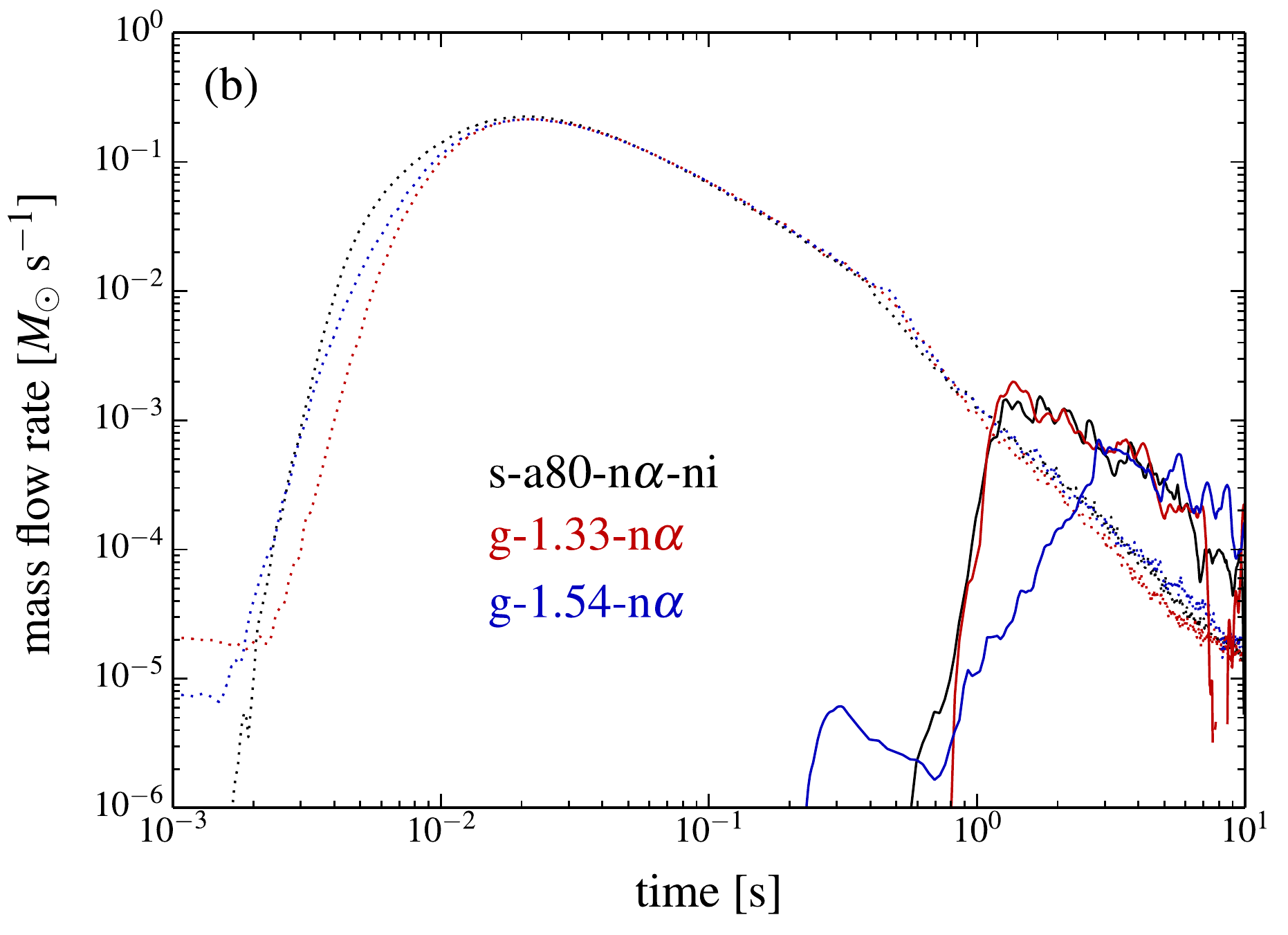}
\label{f:mdot_mout_appendix}
\caption{Comparison of mass outflow and accretion in models with physical
and $\gamma$-law EOSs. Shown is the mass accretion rate at the ISCO (dotted lines)
and outflow rates at $10^9$~cm (solid lines) for selected models from Table~\ref{t:appendix}.
Panel (a) shows the effect of removing key physics from the s-a80 series
of models, while panel (b) compares the $\gamma$-law models with the
physical model with no $\alpha$ particle recombination or neutrino self-irradiation. 
The best agreement in mass ejection is obtained with $\gamma_{\rm ad} = 4/3$.}
\end{figure*}

Here we compare the results from disk wind simulations using a physical
EOS and a $\gamma$-law EOS, to choose the appropriate
value of the adiabatic index for {\tt HARMPI} GRMHD models and their
hydrodynamic counterparts.

As a baseline of comparison, we take a representative disk
model of F15, which includes the EOS
of \citet{timmes2000} with the abundances of neutrons, protons, 
and alpha particles in NSE. We focus on model s-a80, which has 
an initial disk mass $M_{{\rm t}0} = 0.03M_\sun$, a central BH with mass 
$M_{\rm bh}= 3M_\sun$ and spin $a=0.8$, uses a viscosity parameter $\alpha=0.03$,
and employs optically thin neutrino source terms with optical depth corrections of
the form $e^{-\tau}$, with $\tau$ an approximate optical depth.
In addition to the fiducial model s-a80, we evolve three variants that
remove key physics: one without nuclear recombination (s-a80-n$\alpha$), 
one without neutrino self-irradiation (s-a80-ni), and one
without both nuclear recombination and neutrino self-irradiation (s-a80-n$\alpha$-ni).
The grid is similar to that described in \S\ref{s:hydro_methods}, but with 
a resolution of 56 cells in the polar direction and 64 cells per decade 
in radius, with an effective cell size $\Delta r/r \simeq \Delta \theta \simeq 2^\circ$
at the equator (lower in each direction by a factor two relative to the hydrodynamic
models in Table~\ref{t:models}).

We evolve 3 comparison $\gamma$-law models at the same resolution, using the physics 
described in \S\ref{s:initial_model}. Two models exclude nuclear
recombination, one of them using the adiabatic index at the initial density maximum
in model s-a80 ($\gamma_{\rm ad} = 1.54$, model g-1.54-n$\alpha$), and the other using the adiabatic
index in the wind at late times ($\gamma_{\rm ad} = 4/3$, model g-1.33-n$\alpha$). 
A third model includes nuclear recombination as described in \S\ref{s:initial_model}
using $\gamma_{\rm ad}=4/3$, and is thus a lower resolution version of model h2d-v03 from Table~\ref{t:models}.

Table~\ref{t:appendix} shows the
mass ejection properties for the comparison models,
with selected cases shown in Figure~\ref{f:mdot_mout_appendix}. 
The exclusion of
nuclear recombination in the models with a physical EOS causes a
$\sim 30\%$ reduction in the total mass ejection, whereas excluding
self-irradiation hardly changes the result. Excellent agreement
is found between models s-a80-n$\alpha$-ni and g-1.33-n$\alpha$ 
in the integrated and instantaneous mass ejection. In contrast,
the model with $\gamma_{\rm ad}=1.54$ ejects a much smaller amount of mass
($30\%$ less than the model with $\gamma_{\rm ad}=4/3$). On the other hand,
the accretion history at the ISCO is most similar for models
s-a80-n$\alpha$-ni and g-1.54-n$\alpha$. \emph{Given that this study
focuses on the late-time wind properties, we adopt $\gamma_{\rm ad}=4/3$
as the adiabatic index of our fiducial production models}. 

The contribution from alpha particle recombination is
stronger in model g-1.33 than in s-a80-ni. Note however that doubling
the resolution in angle and radius can also lead to differences
in mass ejection of the order of $\sim 10\%$, so we consider agreement
between these two models to be acceptable.

\begin{table}
\centering
\begin{minipage}{8cm}
\caption{Comparison models and mass ejection properties. Columns from left to right
show model name, type of EOS (helm: physical, $\gamma_{\rm ad}$: gamma-law), inclusion of
$\alpha$ particle recombination energy, inclusion of neutrino self-irradiation, 
total mass ejection in units of the initial torus mass, and time- and 
mass-flux-averaged $Y_e$  in the wind.\label{t:appendix}}
\begin{tabular}{lccccc}
\hline
{Model}&
{EOS} &
{$\alpha$} &
{Irr.} &
{$M_{\rm ej}/M_{{\rm t}0}$} &
{$\bar{Y}_e$}\\
\hline
s-a80               & helm          & Y   & Y   & 0.18 & 0.22  \\
s-a80-ni            &               &     & N   & 0.18 & 0.22  \\
s-a80-n$\alpha$     &               & N   & Y   & 0.13 & 0.23  \\
s-a80-n$\alpha$-ni  &               &     & N   & 0.12 & 0.23  \\
\noalign{\smallskip}
g-1.33              & $\gamma_{\rm ad}=4/3$  & Y   & N   & 0.20 & 0.21  \\
g-1.33-n$\alpha$    &               & N   &     & 0.12 & 0.22  \\
g-1.54-n$\alpha$    & $\gamma_{\rm ad}=1.54$ &     &     & 0.09 & 0.20  \\
\hline
\end{tabular}
\end{minipage}
\end{table}

\subsection{Degeneracy}
\label{s:degeneracy}

While the effects of electron degeneracy on the total pressure are
not dominant, its inclusion in the neutrino rates is crucial
for obtaining an adequate level of neutron richness. In particular,
degeneracy strongly suppresses the positron density,
because pairs are created in the exponential tail of the thermal
distribution above the Fermi sphere (e.g., \citealt{Beloborodov03}). 
The lack of positron captures leads to an overabundance of neutrons.

We compute the level of electron degeneracy analytically in the
relativistic limit ($kT \gg m_e c^2 \simeq 0.5$~MeV). We start
by writing the net electron density in terms of the degeneracy parameter (e.g., \citealt{bethe80}):
\begin{eqnarray}
n_{e^-} - n_{e^+} & = & \frac{1}{\pi^2}\left(\frac{kT}{\hbar c} \right)^3
                   \left[\mathcal{F}_2(\eta_e)-\mathcal{F}_2(-\eta_e) \right]\\
                  & = & \frac{1}{3\pi^2}\left(\frac{kT}{\hbar c} \right)^3
                   \left[\pi^2\eta_e + \eta_e^3\right]
\end{eqnarray}
where $\eta_e = \mu_e/(kT)$ is the chemical potential of electrons. 
Writing $n_{e^-}-n_{e^+} = \rho Y_e/m_n$ and using the Fermi momentum 
$p_{\rm F} = \hbar\left(3\pi^2 Y_e\rho/m_n \right)^{1/3}$, we obtain a 
cubic equation for $\eta_e$ in terms of $\left\{\rho, T, Y_e\right\}$:
\begin{equation}
\eta_e^3 + \pi^2\eta_e -\left(\frac{p_{\rm F}c}{kT} \right)^3 = 0.
\end{equation}
This equation has the solution
\begin{eqnarray}
\eta_e & = & \left[\frac{1}{2}\left(\frac{p_{\rm F}c}{kT} \right)^3 
             + \sqrt{\frac{1}{4}\left(\frac{p_{\rm F}c}{kT} \right)^6 + \frac{\pi^6}{27}}\right]^{1/3}\nonumber\\
\label{eq:etae_solution}
       &&  \qquad+ \left[\frac{1}{2}\left(\frac{p_{\rm F}c}{kT} \right)^3
             - \sqrt{\frac{1}{4}\left(\frac{p_{\rm F}c}{kT} \right)^6 + \frac{\pi^6}{27}}\right]^{1/3},
\end{eqnarray}
with limiting cases
\begin{equation}
\eta_e \simeq
\begin{dcases}
\left(\frac{p_{\rm F}c}{kT}\right)                  & \qquad p_{\rm F}c\gg \pi kT\\
\frac{1}{\pi^2}\left(\frac{p_{\rm F}c}{kT}\right)^3 & \qquad p_{\rm F}c\ll \pi kT.
\end{dcases}
\end{equation}
We set $\eta_e = 0$ for $T < 1$~MeV, for which neutrino source terms become sub-dominant.

To include the effects of degeneracy in the neutrino rates, we need to calculate
individual Fermi functions for input in eqns.~(\ref{eq:D4})-(\ref{eq:D5}) and evaluate
them with $\eta_e$ obtained from eq.~(\ref{eq:etae_solution}). The Fermi
functions of negative argument are obtained using a
well-known series expansion (e.g., \citealt{mcdougall1938}):
\begin{eqnarray}
\mathcal{F}_4(-\eta) & = & 4!\left[e^{-\eta}-\frac{e^{-2\eta}}{2^5}+ \frac{e^{-3\eta}}{3^5} - ...\right]\\
\mathcal{F}_5(-\eta) & = & 5!\left[e^{-\eta}-\frac{e^{-2\eta}}{2^6}+ \frac{e^{-3\eta}}{3^6} - ...\right].
\end{eqnarray}
We take the first three terms in the series, as in \citet{shibata2007}, yielding values that
are accurate to better than 1\% for $\eta = 0$.
We then obtain the Fermi functions of positive argument using the exact relations between sums 
and differences from \citet{bludman1978}:
\begin{eqnarray}
\mathcal{F}_4(\eta)-\mathcal{F}_4(-\eta) & = & \frac{7\pi^4}{15}\eta + \frac{2\pi^2}{3}\eta^3
                                           + \frac{1}{5}\eta^5\\
\mathcal{F}_5(\eta)+\mathcal{F}_5(-\eta) & = & \frac{31\pi^6}{126} + \frac{7\pi^4}{6}\eta^2
                                           +\frac{5\pi^2}{6}\eta^4 + \frac{1}{6}\eta^6.
\end{eqnarray}

The comparison models shown in Table~\ref{t:appendix} indicate that including degeneracy 
in the rate of change of $Y_e$ due to neutrino and antineutrino emission (equation~\ref{eq:Gnu}) 
leads to average electron fractions that are very similar to those obtained
when more physics is included (as in the  s-a80-type models).

\subsection{Approximation to Nuclear Statistical Equilibrium}
\label{s:recombination}

\begin{figure}
\includegraphics*[width=\columnwidth]{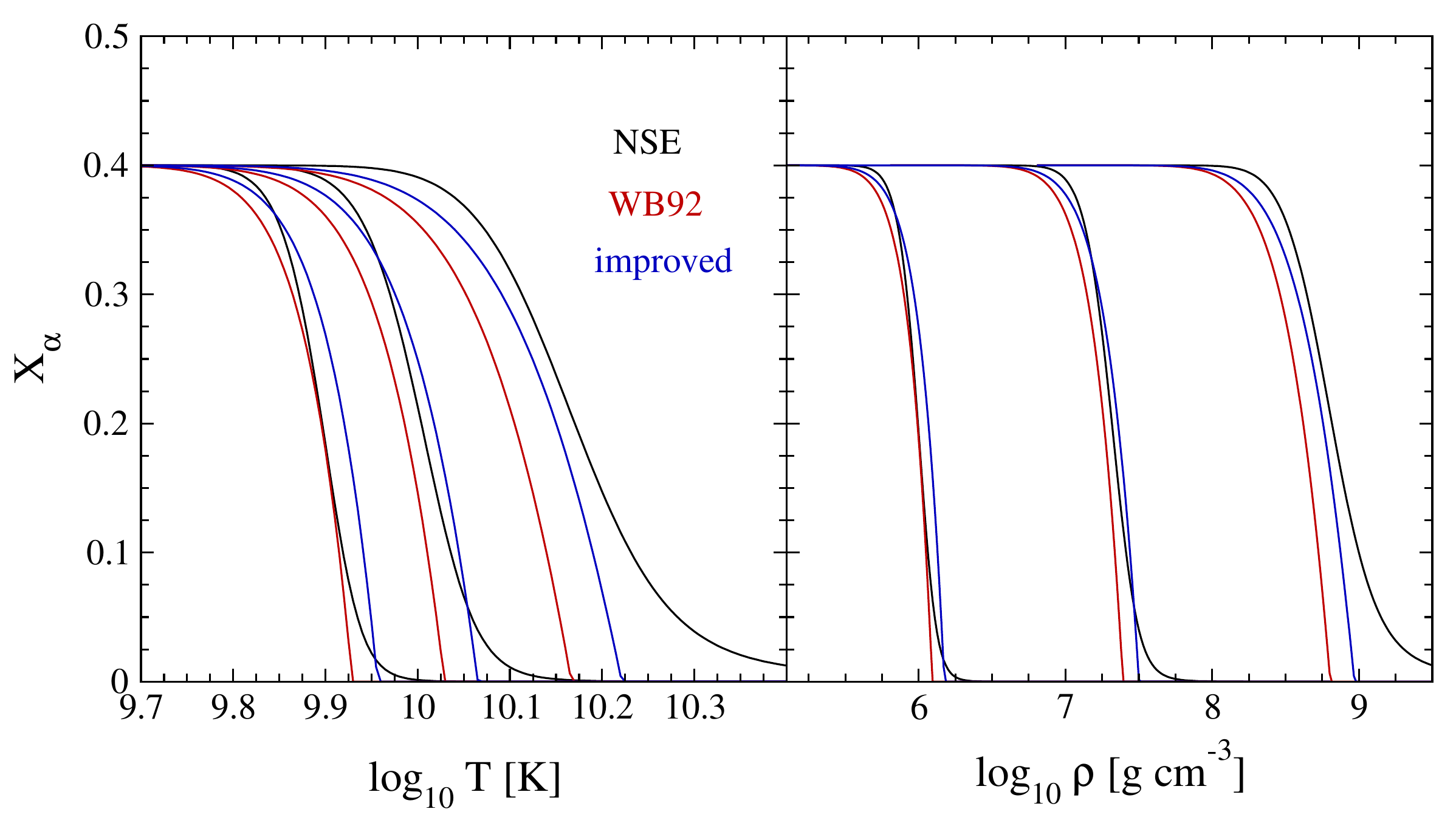}
\caption{Comparison between the mass fractions of alpha particles obtained from an 
NSE calculation (black), the analytic fit of \citet{Woosley&Baron92} (red), and
equation~(\ref{eq:xalpha_fit}) (blue). The latter differs from \citet{Woosley&Baron92}
by a global factor of $2^{-3/4}$ introduced to improve agreement in the thermodynamic
regime of interest. The thermodynamic trajectories assume $T^3/\rho = $ constant 
and $Y_e = 0.2$.}
\label{f:xalpha_fit}
\end{figure}

Figure~\ref{f:xalpha_fit} compares
the results of an NSE calculation for the mass fraction of $\alpha$ particles
(e.g., \citealt{Shapiro&Teukolsky83}) 
with the analytic fit of \citet{Woosley&Baron92}. This fit has been
used in a number of NS merger remnant disk studies to parameterize
the effects of nuclear recombination (e.g., 
\citealt{popham1999,DiMatteo+02,lee2005b,shibata2007}).
The curves in Figure~\ref{f:xalpha_fit} keep $T^3/\rho$
constant, as appropriate for wind particles in a medium dominated by radiation
pressure and in which other energy source terms are unimportant, and assume $Y_e = 0.2$.

Figure~\ref{f:xalpha_fit} shows curves for three values of the entropy which span 
the relevant thermodynamic range spanned by the wind. The fit of \citet{Woosley&Baron92} systematically
underestimates the characteristic densities and temperatures at the point where the transition from
$\alpha$-particles to pure nucleons takes place, with the error being larger at 
higher densities. Nonetheless, a minor adjustment 
to the numerical prefactor, from $26.2$ to $15.6\simeq 26.2/2^{3/4}$, shifts the curves closer to 
the NSE results. Quantitatively, the fractional errors in the temperatures and densities
at the point where $X_\alpha = Y_e = 0.2$ are
$\{5,3,5\}\%$ and $\{16,10,13\}\%$, for the low, medium, and high density trajectories, respectively.

\bibliographystyle{mn2e}
\bibliography{ms,rodrigo}

\label{lastpage}
\end{document}